\pdfoutput=1
\documentclass{article}
\usepackage{caption} 
\usepackage{graphicx}
\usepackage{subfigure}
\usepackage{amsmath}
\usepackage{geometry}
\usepackage{authblk}
\usepackage{hyperref}
\usepackage[T1]{fontenc}
\usepackage[utf8]{inputenc}
\usepackage{float}
\usepackage{amssymb}

\usepackage{tikz}
	
\usetikzlibrary{backgrounds,fit,decorations.pathreplacing}
\newcommand{\ket}[1]{\ensuremath{\left|#1\right\rangle}}

\title{Quantum Cluster Algorithm for Data classification}
\author{Junxu Li and Sabre Kais\thanks{Email: kais@purdue.edu}}
\affil{Department of Chemistry, Department of Physics and Astronomy, 

and Purdue Quantum Science and Engineering Institute,

	Purdue University, West Lafayette, IN 47907, USA}
\begin{document}
\tikzset{meter/.append style={draw, inner sep=10, rectangle, font=\vphantom{A}, minimum width=30, line width=.8,
 path picture={\draw[black] ([shift={(.1,.3)}]path picture bounding box.south west) to[bend left=50] ([shift={(-.1,.3)}]path picture bounding box.south east);\draw[black,-latex] ([shift={(0,.1)}]path picture bounding box.south) -- ([shift={(.3,-.1)}]path picture bounding box.north);}}}
\tikzset{
 cross/.style={path picture={ 
\draw[thick,black](path picture bounding box.north) -- (path picture bounding box.south) (path picture bounding box.west) -- (path picture bounding box.east);
}},
crossx/.style={path picture={ 
\draw[thick,black,inner sep=0pt]
(path picture bounding box.south east) -- (path picture bounding box.north west) (path picture bounding box.south west) -- (path picture bounding box.north east);
}},
circlewc/.style={draw,circle,cross,minimum width=0.3 cm},
circle0/.style={draw,circle,minimum width=0.3 cm},
}
	\maketitle
	
	\begin{abstract}
	
	We present a  quantum algorithm for data classification  based on the nearest-neighbor learning algorithm. The classification algorithm is divided into two steps: Firstly, data in the same class is divided into smaller groups with sublabels assisting building boundaries between data with different labels. Secondly we construct a quantum circuit for classification that contains multi control gates. The algorithm is  easy to implement and efficient in predicting the labels of test data. To illustrate the power and efficiency  of this approach, we construct the phase transition  diagram for the metal-insulator transition of $VO_2$, using limited trained experimental data, where $VO_2$ is a typical strongly correlated electron materials, and the metallic-insulating phase transition has drawn much attention in condensed matter physics. Moreover, we demonstrate our algorithm on the classification of randomly generated data and the classification of entanglement for various Werner states, where the training sets can not be divided by a single curve, instead, more than one curves are required to separate them apart perfectly. 
	Our preliminary result shows considerable potential for various classification problems, particularly for constructing  different phases in materials.
	\end{abstract}

\section*{Introduction}
\label{Introduction}

Machine learning techniques have demonstrated remarkable success
in numerous topics in science and engineering, including artificial intelligence \cite{mitchell1990machine,duda2012pattern},  molecular dynamics \cite{botu2015adaptive}, light harvesting systems \cite{hase2017machine}, molecular electronic properties \cite{montavon2013machine}, surface reaction network \cite{ulissi2017address}, density functional models \cite{brockherde2017bypassing}, phase classification, and quantum simulations \cite{wang2016discovering,carrasquilla2017machine,broecker2017machine,ch2017machine,van2017learning,arsenault2014machine,kusne2014fly}. In addition, modern machine learning techniques have also been applied to the state space of complex condensed-matter systems for their abilities to analyze exponentially large data sets \cite{carrasquilla2017machine},  speed-up searches for novel energy generation/storage materials \cite{de2017use,wei2016neural} and classification of  entanglement\cite{gao2018experimental}. With the rapid development of quantum computers\cite{leibfried2003experimental,debnath2016demonstration, karra2016prospects, arute2019quantum, zhong2020quantum}, it has become a new frontier to recognize patterns using quantum computers. Considering recent advancements in both quantum computing and machine learning, the combination of the two techniques – quantum machine learning – is expected to be a promising application of quantum computer in the near future. Many quantum machine learning algorithms were proposed in the past few years\cite{rebentrost2018quantum,cao2016combined,biamonte2017quantum,roy2021enhancement,dixit2021training}. 
Moreover, researchers have succeeded to apply quantum machine learning  algorithms to  various systems such as superconducting circuits\cite{havlivcek2019supervised} and  photonic systems\cite{cai2015entanglement},
which leads to enormous enthusiasm applying quantum algorithms into various areas\cite{xia2018quantum,hu2020quantum,hu2020quantum,li2021practical,xia2017electronic,sajjan2021quantum,xia2021quantum}.
	
There is no doubt that we are now in the age of big data and there is an urgent need for developing game-changing quantum algorithms to perform machine learning tasks on large-scale scientific datasets for various industrial and technological applications based on optimization. For a proof of concept,  Du and coworkers have successfully distinguished handwriting numbers '6' and '9' with the quantum support vector machine \cite{li2015experimental}. However, it could be difficult to deal with more challenging problems
, especially when the training data can not be divided apart by a single curve, instead, more than one curve or even enclosed curves might be required to separate them apart.
Another remarkable development  is applying  quantum machine learning   on  variational circuits\cite{schuld2019quantum,arrazola2018machine}, which theoretically, can always be able to classify data with complex distribution. Yet generally, these algorithms rely heavily on a gradient-based systematic optimization of parameters\cite{mitarai2018quantum,farhi2018classification}. On the other hand, quantum nearest neighbor algorithm\cite{wiebe2014quantum} offers  another option to classify data without the gradient based optimization process. In brief, the core of nearest-neighbor classification algorithm is to assign the training vectors into classes, and in each class vectors are close to each other.

	In this study, we will propose a quantum classification algorithm, with which we can build a quantum circuit that is able to classify artificially generated data, and all parameters in the circuit can be obtained without relying on the gradient based optimization process. For this purpose, we introduce 'sublabels' to assist in  classifying  data with intricate distribution, where 'sublabel' represents a minor label subordinates to the main one,  it also called 'subclass'. There are two main tasks in our developed algorithm: how to find the appropriate sublabels and how to build the  quantum classification circuit with these sublabels.
	With the numerical simulation we will demonstrate the application on various classification problems, especially on constructing different phases of materials.
	First, in section 1,  we will present  the basic elements of the algorithm.  Then, we will apply the  algorithm  for classifications for several systems: classification of metallic and insulating phases in the phase diagrame of  VO$_2$;
classification of entanglement in Werner states, and  classification of randomly generated data.
Finally, we will present scaling analysis and discuss generalization  of  quantum classification algorithm in higher dimensional space.  
In addition, we present in the supplementary materials all the details of the quantum classification algorithm with examples.

	\section{Algorithm design}
	\label{Algorithm design}
	
	Consider the training data set $\{{\bf x}_i, y_i\}$, where ${\bf x}_i$ is a vector in $\mathbb{R}^d$, where $d$ is the dimension and $y_i$ represents the label with possible values $\{l_1, l_2, \dots l_{M}\}$. Our goal is to build a quantum classification circuit that can be used to  predict the label for new vectors $\{\bf x_t\}$. 
	The classification algorithm is divided into two steps:  The first step is a learning process, where one needs to find the “sublabels” for each class of the training data. Then, based on the information obtained in this learning process, we construct a quantum classification circuit that contains multi control quantum gates.

	In the learning process, firstly we apply the Lloyd's algorithm\cite{mackay2003information} for unsupervised machine learning, which assigns training vectors to the same class as the closest mean vector. However the results derived by Lloyd's algorithm can not be used directly as there might be a sublabel redundancy or not  enough sublabels to reconstruct the initial distribution. To address this issue, in addition to the algorithm for clustering, we propose to use two adjusting algorithms:  one to reduce excessive sublabel and  the other to  make sure that there is no overlap between sublabels.
	
\begin{figure}[H]
		\begin{center}
			\centerline{
    \begin{tikzpicture}[thick]
    %
    \tikzstyle{operator} = [draw,fill=white,minimum size=1.5em] 
    \tikzstyle{phase} = [fill,shape=circle,minimum size=5pt,inner sep=0pt]
    \tikzstyle{surround} = [fill=blue!10,thick,draw=black,rounded corners=2mm]
    %
    \node at (0,0) (q1) {\ket{0}};
    %
    \node[operator] (op11) at (2,0) {$R_z(\frac{\phi_1}{2})$} edge [-] (q1);
    %
    \node[operator] (op12) at (4,0) {$R_y({-\theta_1})$} edge [-] (op11);
    %
    \node[operator] (op13) at (6,0) {$R_z(-\frac{\phi_1}{2})$} edge [-] (op12);
    %
    \node[operator] (op14) at (8,0) {$R_z(\frac{\phi_2}{2})$} edge [-] (op13);
    %
    \node[operator] (op15) at (10,0) {$R_y({\theta_2})$} edge [-] (op14);
    %
    \node[operator] (op16) at (12,0) {$R_z(-\frac{\phi_2}{2})$} edge [-] (op15);
    \node (end1) at (13.5,0) {} edge [-] (op16);
    \begin{pgfonlayer}{background} 
    \node[surround] (background) [fit = (q1) (end1)] {};
    \end{pgfonlayer}
    \end{tikzpicture}
  }
	    \captionsetup{justification=centering}
	    \caption{
				{\bf Sketch of the quantum circuit for estimating the inner product:} Circuit to estimate the  inner product of two-dimensional  vectors contains six rotational  gates. Additionally, we also need a memory to store the information of this group, which could be represented by $\theta_{m}$, $\phi_{m}$, and an integer $N$, which represents the total number of data in this group. }
		\label{device}
		\end{center} 
\end{figure}
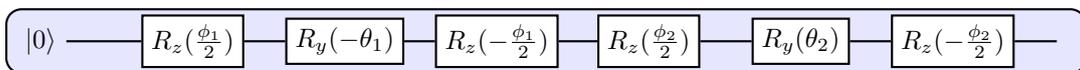

	For each sublabel, we need to store the  information and build a quantum circuit to estimate the inner product, which is shown in fig.(\ref{device}). When ${\bf x}_i$ are vectors in two-dimensional  space, each data point (or sublable) could be represented by a single qubit.
	An arbitrary state of a single qubit could be written as
	
	\begin{equation}
	    |\psi\rangle = e^{i\alpha}[\ cos(\theta/2)|0\rangle + sin(\theta/2)e^{-i\phi/2}|1\rangle\ ]
	    \label{1qbstate} 
	\end{equation}
	
	where $\alpha$ is the global phase and the vector ${\bf x}_i= (x_{1,i},x_{2,i})$ is mapped as
	\begin{align*}
	    \theta_i = \frac{2\pi( x_{1,i}-min\{x_1\})}{max\{x_1\}-min\{x_1\}}\\
	    \phi_i = \frac{2\pi (x_{2,i}-min\{x_2\})}{max\{x_2\}-min\{x_2\}}
	\end{align*}
	Here $max\{x_1\}$ and $max\{x_2\}$  represent the maximum value of all $x_{1,i}$  and  $x_{2,i}$ respectively, while $min\{x_1\}$ and $min\{x_2\}$  represent the minimum value of all $x_{1,i}$  and  $x_{2,i}$ respectively.  Then we need to find a measure  describing the distance between the two states, where the 'distance' might be the Euclidean distance between the two vectors\cite{wiebe2014quantum}, or the inner product of their two corresponding quantum states. Here, we chose to calculate the inner product, as calculating the Euclidean distance is more time  and resource-consuming.
 
	An arbitrary state of a single qubit as shown in Fig.(\ref{1qbstate}),  could be prepared by three rotational  gates:
	\begin{equation}
	    |\psi(\theta_1, \phi_1)\rangle = R_z(\phi_1/2)R_y(\theta_1)R_z(-\phi_1/2)\ |0\rangle
	\end{equation}
	Thus, the inner product between $|\psi(\theta_1, \phi_1)\rangle$ and $|\psi(\theta_2, \phi_2)\rangle$ is given by:
	\begin{equation}
	    \langle\psi(\theta_1, \phi_1)|\psi(\theta_2, \phi_2)\rangle
	    =\langle 0|R_z(\phi_1/2)R_y(-\theta_1)R_z(-\phi_1/2)R_z(\phi_2/2)R_y(\theta_2)R_z(-\phi_2/2)|0\rangle .
	\end{equation}
	The circuit that estimates the inner product will contain six rotational gates, as shown in Fig.(\ref{device}). After a measurement of the final state in the $Z$-basis, the probability of  getting a state  $|0\rangle$ will be an estimation of the inner product $\langle\psi(\theta_1, \phi_1)|\psi(\theta_2, \phi_2)\rangle$. For simplicity, in the following sections, we will write 
	\begin{equation*}
	    R(\theta, \phi)=R_z(+\phi/2)R_y(\theta)R_z(-\phi/2)
	\end{equation*}
	
Moreover,  for every sublabel it is required to store 2 floating  numbers $\theta_m$ and $\phi_m$ that represent the centroid vector of this subgroup, and an integer N representing the total number of data points in this subgroup. 
	
	In the learning process, three  basic algorithms are applied to assist in obtaining “sublabels” from the given training data. Algorithm (\ref{algorithm1}) is designed for an initial  clustering of  the training data. When designing Algorithm (\ref{algorithm1}),  we refer  to Lloyd's algorithm\cite{mackay2003information}in which  we need to assign each vector to the cluster with the closest mean, and then recalculate the centroids of the new cluster. Algorithm (\ref{algorithm1}) will divide the training data with the same prior label into several subgroups.  Algorithm (\ref{algorithm2}) will reduce redundancy, and Algorithm (\ref{algorithm3}) is introduced to make sure there will be no overlap between any two left sublabels of the different prior labels. 
    The goal is to leave only the minimal sublabels without losing important information of the training data. After applying Algorithm (\ref{algorithm1}), and repeating Algorithm (\ref{algorithm2}), Algorithm (\ref{algorithm3}) for a number of times, we can get a set with minimal sublabels and information of the centroid vectors for each subgroup (Details of these three algorithms are in the supplementary materials).
	
	Now, the next step is to build the quantum classification circuit based on the previously obtained information. Consider the following sublabel-control operations,
	
	\begin{eqnarray*}
	\left.\begin{aligned}
	    &\ {\bf IF} \ SUB LABEL = l_j\\
	    &\ \qquad{\bf DO} \quad ROTATION \ R(-\theta^{l_j}_{m}, -\phi^{l_j}_{m})\\
	    &\ {\bf IF} \ SUB LABEL = l_{j+1}\\
	    &\ \qquad{\bf DO} \quad ROTATION \ R(-\theta^{l_{j+1}}_{m}, -\phi^{l_{j+1}}_{m})\\
	    &\ \cdots\\
	    &\ {\bf FOR}\  SUB LABEL \ in\  PRIOR LABEL \ L_i\\
	    &\qquad{\bf DO} \quad OPERATION \ U(L_i)
	    \end{aligned}\right.
	\end{eqnarray*}
	
	where the operations $U(L_i)$ are obtained with the  aim to reach the final state $|\Psi^{f}_{L_i}\rangle$. If one wants to set all vectors belong to prior label $L_i$ close to the final state $|\Psi^{f}_{L_i}\rangle$, then $U(L_i)$ can be chosen to satisfy
	\begin{equation*}
	    |\Psi^{f}_{L_i}\rangle = U(L_i)|0\rangle
	\end{equation*}

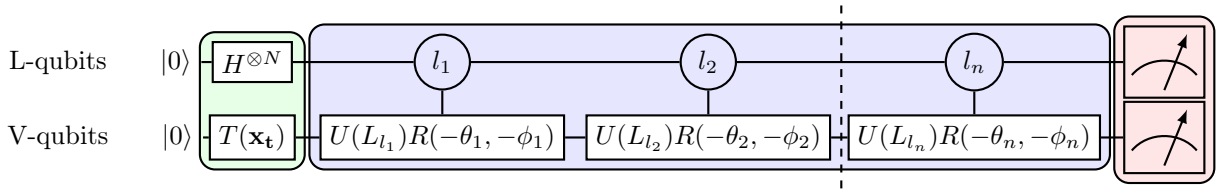
\begin{figure}[H]
			\centerline{
    \begin{tikzpicture}[thick]
    \tikzstyle{operator} = [draw,fill=white,minimum size=1.5em] 
    \tikzstyle{phase} = [fill,shape=circle,minimum size=5pt,inner sep=0pt]
    \tikzstyle{surround} = [fill=blue!10,thick,draw=black,rounded corners=2mm]
    \tikzstyle{sd_blue} = [fill=blue!10,thick,draw=black,rounded corners=2mm]
    \tikzstyle{sd_red} = [fill=red!10,thick,draw=black,rounded corners=2mm]
    \tikzstyle{sd_yellow} = [fill=yellow!10,thick,draw=black,rounded corners=2mm]
    \tikzstyle{sd_green} = [fill=green!10,thick,draw=black,rounded corners=2mm]
    %
    \node at (0,0) (q1) {L-qubits\qquad\ket{0}};
    \node at (0,-1) (q2) {V-qubits\qquad\ket{0}};
    \node[operator] (op11) at (2,0) {$H^{\otimes N}$} edge [-] (q1);
    \node[operator] (op21) at (2,-1) {$T(\bf x_t)$} edge [-] (q2);
    \node[circle0] (op12) at (4.5,0) {$l_1$} edge [-] (op11);
    \node[operator] (op22) at (4.5,-1) {$U(L_{l_1})R(-\theta_1,{-\phi_1})$} edge [-] (op21);
    \draw[-] (op12) -- (op22);
    \node[] (achor0) at (3,0.25) {};
    \node[circle0] (op13) at (8,0) {$l_2$} edge [-] (op12);
    \node[operator] (op23) at (8,-1) {{$U(L_{l_2})R(-\theta_2,{-\phi_2})$}} edge [-] (op22);
    \draw[-] (op13) -- (op23);
    \node[circle0] (op14) at (11.5,0) {$l_n$} edge [-] (op13);
    \node[operator] (op24) at (11.5,-1) {{$U(L_{l_n})R(-\theta_n,{-\phi_n})$}} edge [-] (op23);
    \draw[-] (op14) -- (op24);
    \node[meter] (meter1) at (14,0) {} edge [-] (op14);
    \node[meter] (meter2) at (14,-1) {} edge [-] (op24);
    \draw [dashed] (9.75,0.75) -- (9.75,-1.75);
    \begin{pgfonlayer}{background} 
    \node[sd_green] (background) [fit = (op11) (op21)] {};
    \node[sd_blue] (background) [fit = (achor0) (op24)] {};
    \node[sd_red] (background) [fit = (meter1) (meter2)] {};
    \end{pgfonlayer}
    \end{tikzpicture}
  }
	    \captionsetup{justification=centering}
	    \caption{
				{\bf Sketch of the main circuit :} Qubits in the main circuit can be divided into two groups: L-qubits to represent the sub labels, and will play the role of control bits and the  V-qubits to represent a  given vector. Initially, L-qubits will be prepared at state $(\frac{|0\rangle+|1\rangle}{\sqrt{2}})^{\otimes N}$, where there are N qubits representing the sub labels. The minimum $N = \lceil\log_2 n\rceil$, and n is the total number of sub labels. In sum $n$ control rotation gates are needed.}
		\label{device2}
	\end{figure}

	To classify a test vector with unknown label, we prefer to rely on a single quantum classification circuit, instead of repeating comparing inner products with the training data. The very basic and intuitive idea is to measure inner product between the new one and all centroid vectors of subgroups. Circuits for classification will consist of two parts: the control qubits representing the sublabels, and others representing the given new vector. Fig.(\ref{device2}) is a sketch showing the structure of the main circuit. First, one map the test data ${\bf x_t}$ into the prepared circuit as 
	\begin{equation}
	    {\bf x_t}\rightarrow|\Psi({\bf x_t})\rangle=\frac{1}{\sqrt{n}}\left[\sum_i^n|l_i\rangle\right]\bigotimes|\psi(\theta,\phi)\rangle
	\end{equation}
	where 
	$|l_i\rangle$ are the orthogonal eigenstates for the control qubits, representing sub labels $l_i$, and there are $n$ sublabels in total. In the mapping process, we need to apply Hadamard gates for L-qubits (representing the sublabels), and apply operator $T({\bf x_t})$ on the V-qubits (representing the test data), where $|\psi(\theta,\phi)\rangle=T({\bf x_t})|0\rangle$. The quantum classification circuit can be described as: 
	\begin{equation}
	    U = \sum_i^n\left[|l_i\rangle\langle l_i|\bigotimes U^{}(L_i) R(-\theta^{l_i}_{m}, -\phi^{l_i}_{m})\right] + \sum_{j=n+1}^{2^N-1}\left[|j\rangle\langle j|\bigotimes I\right]
	\end{equation}
	where there are $N$ qubits used to represent sublabels, and $n$ sublabels totally. We can notice that for $0\leq k\leq n$,
	\begin{equation}
	    \begin{split}
	        \langle l_k, \Psi^{f}_{L_k}| U|\Psi({\bf x})\rangle
	        &=\langle l_k, \Psi^{f}_{L_k}|\sum_i^n
	        \left[|l_i\rangle\bigotimes U^{}(L_i) R(-\theta^{l_i}_{m}, -\phi^{l_i}_{m})|\psi(\theta,\phi)\rangle\right]
	        +\langle l_i, \Psi^{f}_{L_k}|\sum_{j=n+1}^{2^N-1}\left[|j\rangle\langle j|\bigotimes I\right]
	        \\
	        &=\sum_i^n\left[
	        \delta_{ik} \langle\Psi^{f}_{L_i}|U^{}(L_i) R(-\theta^{l_i}_{m}, -\phi^{l_i}_{m})|\psi(\theta,\phi)\rangle
	        \right]
	        \\
	        &=
	        \langle\psi(-\theta^{l_k}_{m}, -\phi^{l_k}_{m})|\psi(\theta,\phi)\rangle
	    \end{split}
	\end{equation}
	By applying a measurement with eigenstate $|l_k\rangle|\Psi^{f}_{L_k}\rangle$, the inner product $\langle\psi(-\theta^{l_k}_{m}, -\phi^{l_k}_{m})|\psi(\theta,\phi)\rangle$ can be estimated.
	Fig.(\ref{device2}) is a sketch of the main circuit, where $T({\bf x_t})$ represents the given test data, $l$ is the sublabel and $L$ is prior label. 
	
	When predicting the label for a new test vector, we assume that the probability for each possible sublabels is the same. Based on this assumption, we applied $N$ Hadamard gates on the label qubits, preparing them as a uniform superposition state. In fact, one can always adjust the probability for sublabels based on the training result, and only keep the states representing 'valid' sublabels. Assume that we finally derived $K$ sublabels, and $N$ qubits are used as label qubits, where $2^{N-1}<K\leq 2^N$.
	For a training data set with two labels $\{{\bf x_i}, y_i\}$, $y_i\in\{0, 1\}$, $K_0$ sublabels are assigned with label $y=0$, while the other $K_1$ labels with $y=1$, $K_0+K_1=K$.
	Then the Hadamard gates $H^{\otimes N}$ can be replaced as some certain operation, in order to prepare the label qubits at state
	\begin{equation}
	    |\phi_L\rangle=
	    \sum_{n=0}^{K_0-1}\sqrt{p_n}|n\rangle + 
	    \sum_{m=0}^{K_1-1}\sqrt{p_{m+K_0}}|m+2^{N-1}\rangle
	    \label{lqubits}
	\end{equation}
	where $p_n\geq0$, and $\sum_{n=0}^{K-1}p_n=1$. Obviously, for sublabels assigned with label $y=0$, the first label qubit is set as state $|0\rangle$, and for the others the first qubit is set as $|1\rangle$. For a new test data, one only needs to measure the first label qubit and the data qubits. By comparing $P(q_1=|y\rangle, V=|\Psi^{f}_{y}\rangle)$, $y_i\in\{0, 1\}$, the new test vector will be assigned with label corresponding to the larger $P$, where $P(q_1=|y\rangle, V=|\Psi^{f}_{y}\rangle)$ represents probability to find the first label qubit $q_1$ at state $|y\rangle$ and meanwhile find the data qubits at state $|\Psi^{f}_{y}\rangle$. $p_n$ are chosen to maximize
	\begin{equation}
	    \sum_i\left\{
	    P(q_1=|y_i\rangle, V=|\Psi^{f}_{y_i}\rangle)
	    -\lambda P(q_1\neq|y_i\rangle, V=|\Psi^{f}_{y_i}\rangle)
	    \right\}
	\end{equation}
	where $\lambda\geq0$ is the penalty.


	\section{Applications}
	\label{Applications}
	\subsection{Classification of metallic-insulating  phases of  vanadium dioxide}
	\label{Classification of metallic-insulating  phases of  vanadium dioxide}
	Strongly correlated electron materials and their phase transitions have attracted great interest for device application and  in condensed matter physics\cite{dagotto2005complexity,arko1997strongly}. Recently, metallic-insulating phase transition of vanadium dioxide ($VO_2$), as a prototype of strongly correlated electronic materials, attracted experimental and theoretical attention for its distinct structures and electronic properties \cite{qazilbash2007mott,jeong2013suppression}. 

\begin{figure}[H]

\centering 

\subfigure[]{
\centering
\label{subfig_a_vo2}
\includegraphics[width=0.45\linewidth]{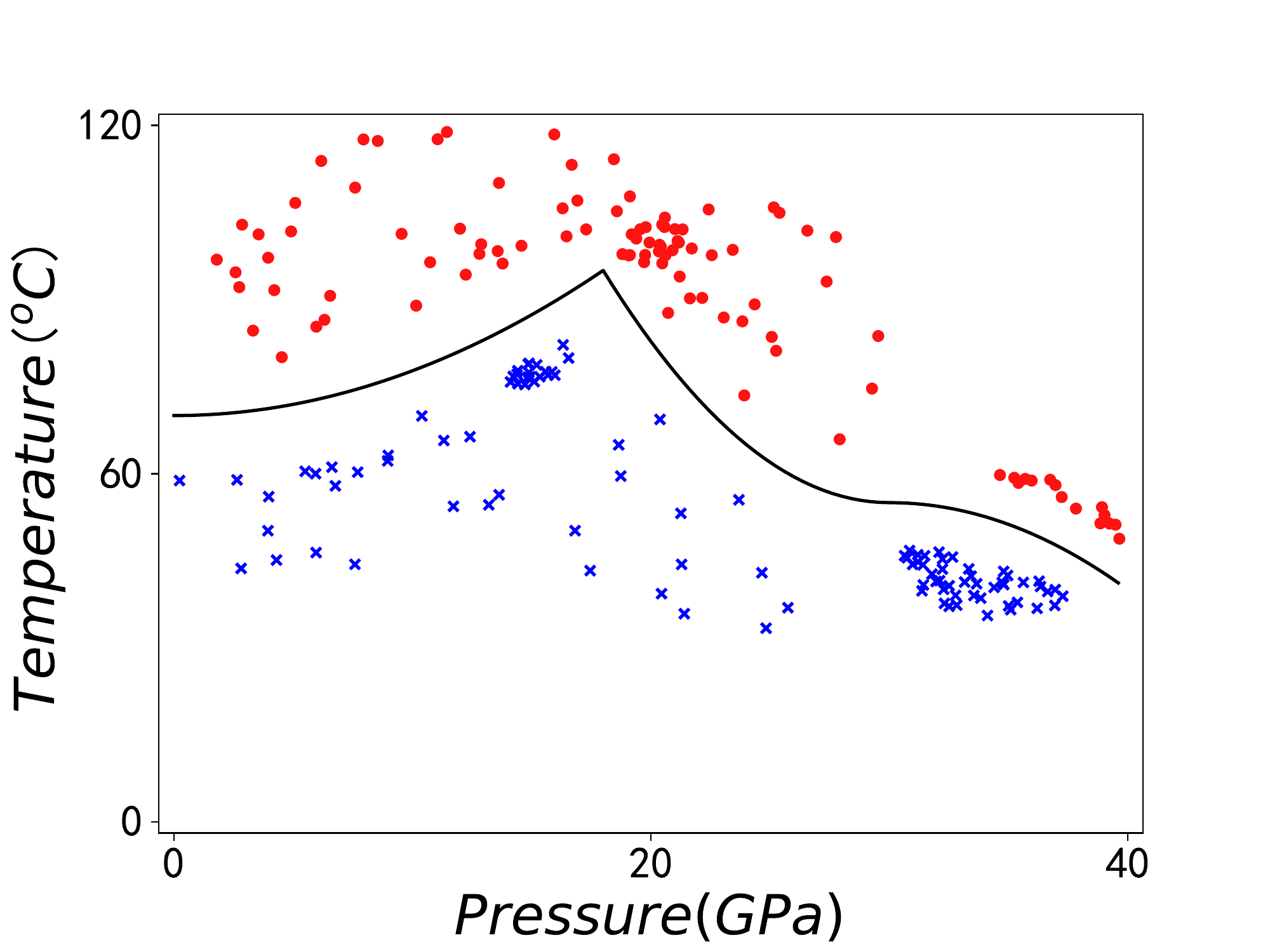}
}
\subfigure[]{
\centering
\label{subfig_b_vo2}
\includegraphics[width=0.45\linewidth]{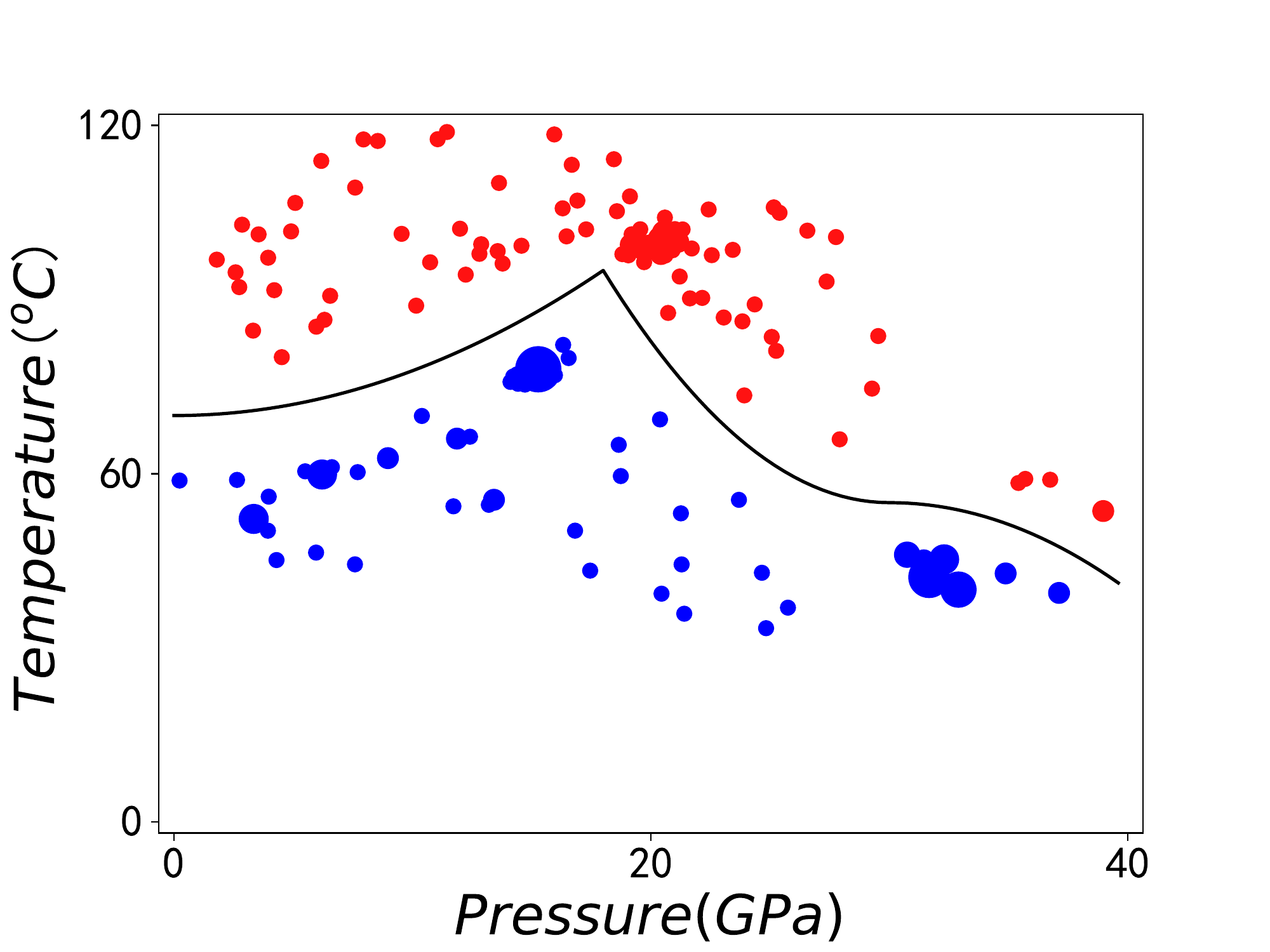}
}

\vfill

\subfigure[]{
\label{subfig_c_vo2}
\includegraphics[width=0.45\linewidth]{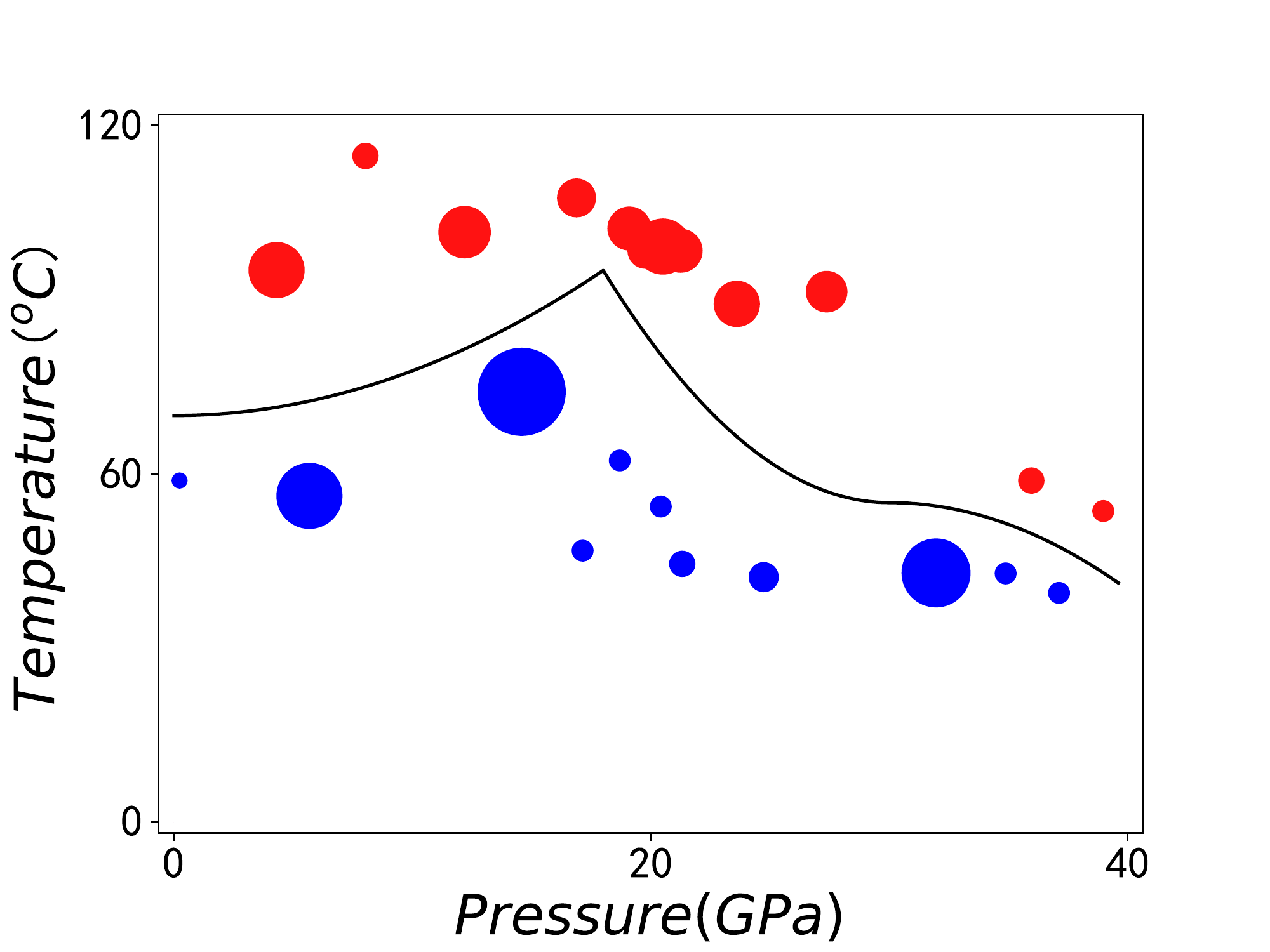}
}
\hspace{0.01\linewidth}
\subfigure[]{
\label{subfig_d_vo2}
\includegraphics[width=0.45\linewidth]{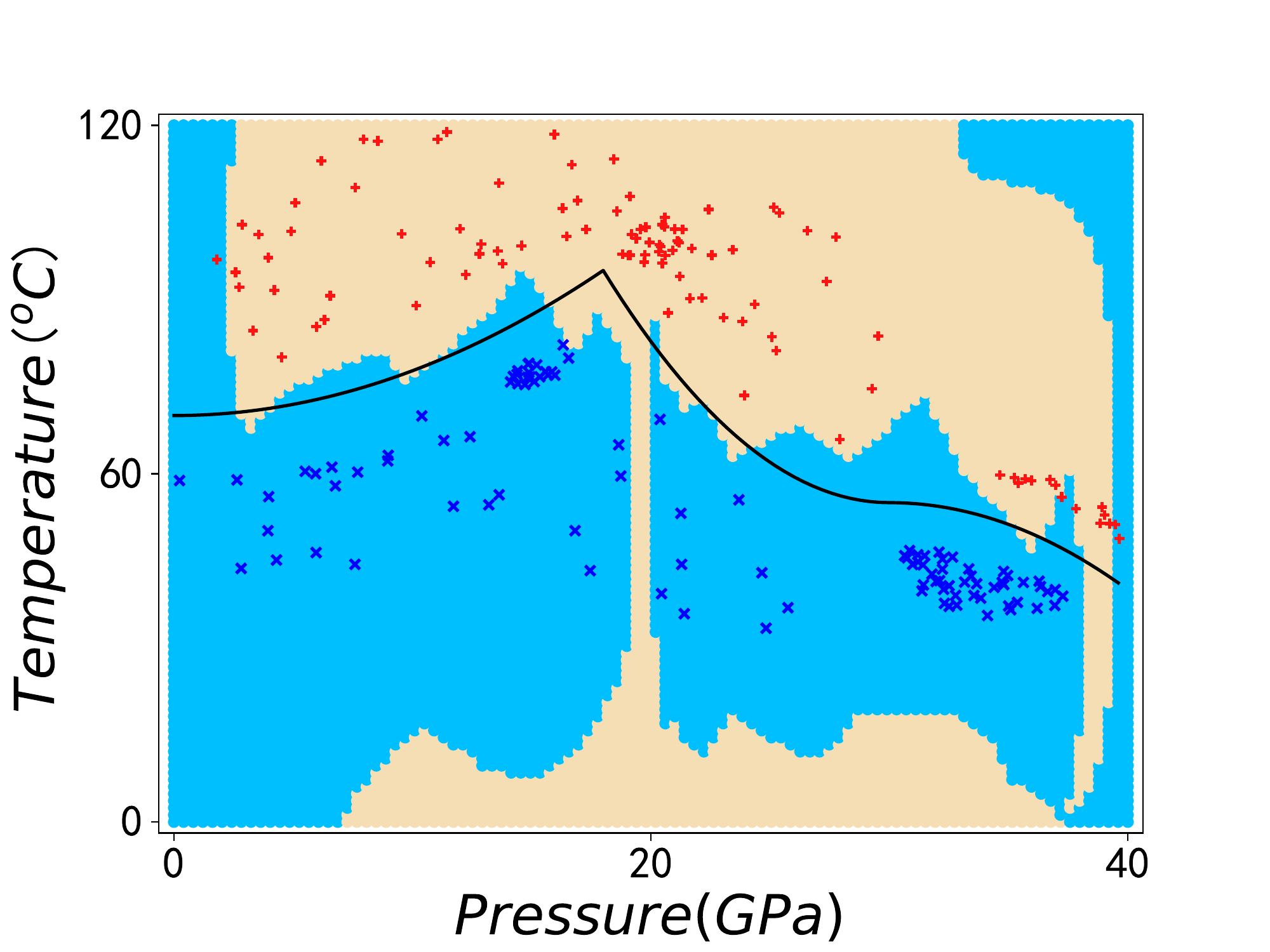}
}
\caption{{\bf Classification of  metallic  and insulating  states of $VO_2$}: (a)Initial data used for classification. Red dots represent metallic state, and blue ones represent insulating state. Phase transition line indicated by  the black solid curve.(b) Forming subgroups after applying algorithm[\ref{algorithm1}] once. Similarly in (b) and (c), blue or red spheres are used to represent data with the same sublabel, where the center of sphere represents the average vector, and the radius represents number of vectors belong to this sublabel. (c) Results after repeating algorithm[\ref{algorithm2},\ref{algorithm3}] 3 times. (d)Prediction of new data. New vector in the blue part will be recognized with label 'insulating', and label of new vectors in yellow part will be predicted as 'metallic'. Blue and red dots are still the initial data.}
\label{fig_simu_vo2}
\end{figure}

Here, we will apply the developed quantum classification algorithm to distinguish the metallic state from the insulating  state of $VO_2$. Data used in this section is based on experimental results reported in Ref. \cite{chen2017pressure}.  $VO_2$ exhibits several special structures under different temperature and pressure. As shown in Fig.(\ref{subfig_a_vo2}), the red dots represent metallic state, the blue dots represent the insulating state, and the black solid line represents the phase transition line. Note, our training data were chosen far from the phase transition line in order to test the classification power of the designed quantum algorithm.  Fig.(\ref{subfig_b_vo2}) show the initial clustering results after applying the Algorithm (\ref{algorithm1}) once. 	In Algorithm (\ref{algorithm1}) one should determine the parameter $D\in(0,1]$, which is the minimum inner product between the centroid vector and arbitrary vectors with the same sublabel. In other words, a vector can be assigned to a certain sublabel when the inner product between itself and the centroid vector of the sublabel are larger than $D$.  Here we set $D=0.99$. Now, we  need to repeat Algorithm (\ref{algorithm2}) and Algorithm (\ref{algorithm3}) several times to reduce the excessive sublabels as shown in Fig.(\ref{subfig_c_vo2}).  After repeating Algorithm(\ref{algorithm2}) and Algorithm (\ref{algorithm3}) three times, the number of classes can not be reduced further as shown
in Fig.(\ref{subfig_c_vo2}).

In both Fig.(\ref{subfig_b_vo2}) and Fig.(\ref{subfig_c_vo2}),  blue and red spheres are used to represent data with the same sublabel, where the center of sphere represents the average vector, and their radius is used to represent the number of vectors belonging  to a  certain sublabel. In Fig.(\ref{subfig_d_vo2}), we demonstrate the prediction results of arbitrary given data. States in the yellow parts are predicted to behave  as metallic states, and the  ones in the blue parts are predicted to be  insulating states. In the training process, we used 100 data vectors for the metallic states and 100 data vectors for the insulating states, out of the totally 1100 data vectors. 
	
	Simulation results shows that our developed quantum algorithm can efficiently classify metallic or insulating  states of $VO_2$.   Finally we need 7 sublabels for insulating  states and 8 sublabels for metallic states, which means that the classification circuit consists of 5 qubits (4 for sublabels as control qubit and 1 for data). 
	
It is important to note,  when Pressure ($P$) or  Temperature ($T$) is small, prediction can be incorrect. The error appears because of the fact that few vectors in these area where  used in the training process. Moreover, when we convert a vector $(P, T)$ into quantum states, we changed them as $\theta, \phi\in[0,2\pi)$. Though classically $T= 0{}^oC$ and $T= 120{}^oC$ are extremely different, when we convert them as angles, $\theta=0$ and $\theta=2\pi$ are nearly the same. One might notice that there is a slim yellow line around $P=20GPa$ in our prediction. It is reported in Ref. \cite{chen2017pressure} that there is a structural transition  between the state $M1$ and $M1'$. However, this is not a metallic-insulating  transition. In our classification algorithm, we did not expect to predict this transition and the slim line shows up “coincidentally”. To get a better prediction results, an option is to map both training data and test data into $[0, \pi]\times[0, \pi]$ instead of $[0, 2\pi]\times[0, 2\pi]$. However, here we do not focus on the  performance at low temperature, nor the phase transition about $P=20GPa$. As a result, mapping data into $[0, 2\pi]\times[0, 2\pi]$ is still an acceptable choice. More simulation results of the phase transition with different training data are offered in the supplementary materials.

    \subsection{Classification of randomly  generated data}
    \label{Classification of randomly  generated data}
    
\begin{figure}[H]

\centering 

\subfigure[]{
\centering
\label{subfig_a_complex}
\includegraphics[width=0.31\linewidth]{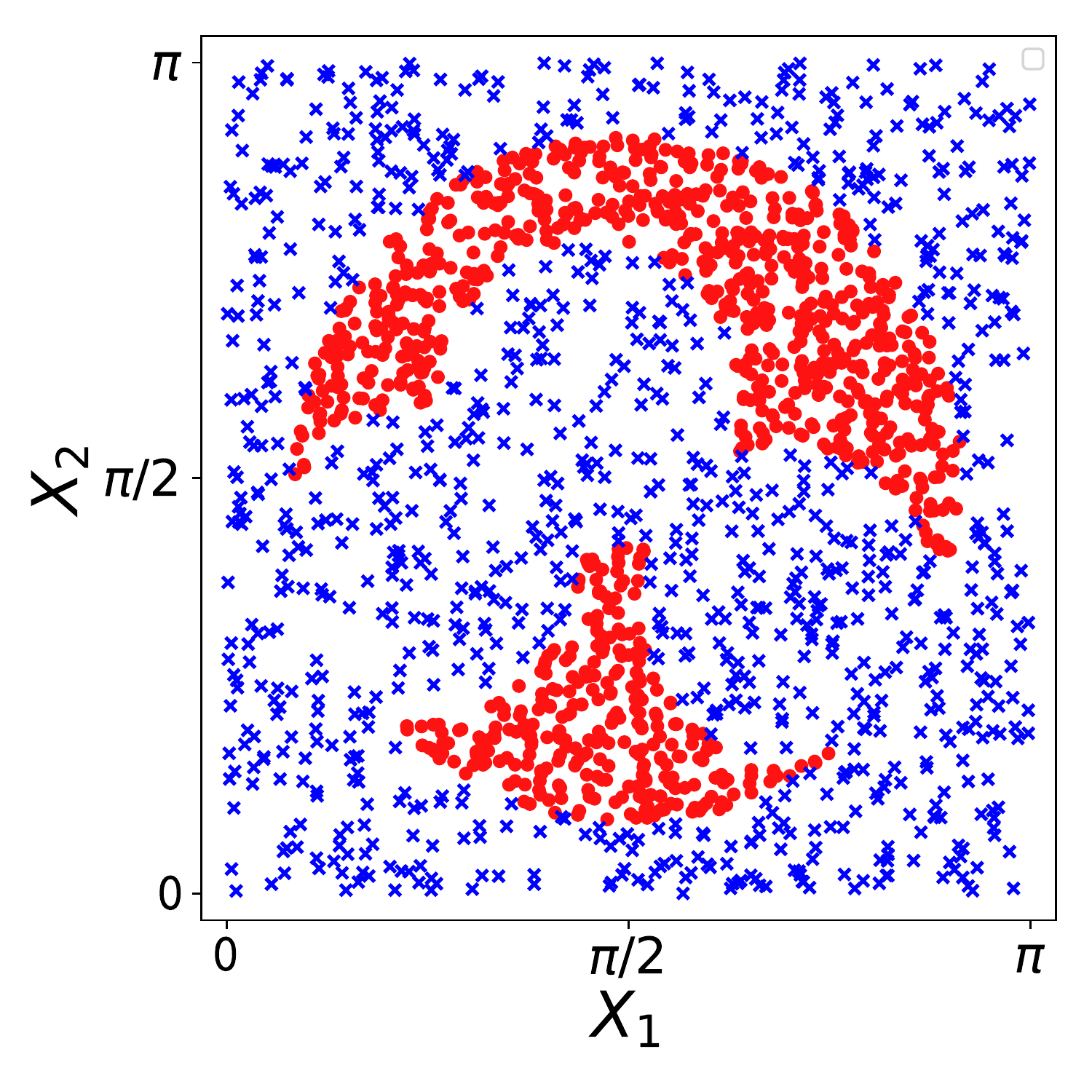}
}
\subfigure[]{
\centering
\label{subfig_b_complex}
\includegraphics[width=0.31\linewidth]{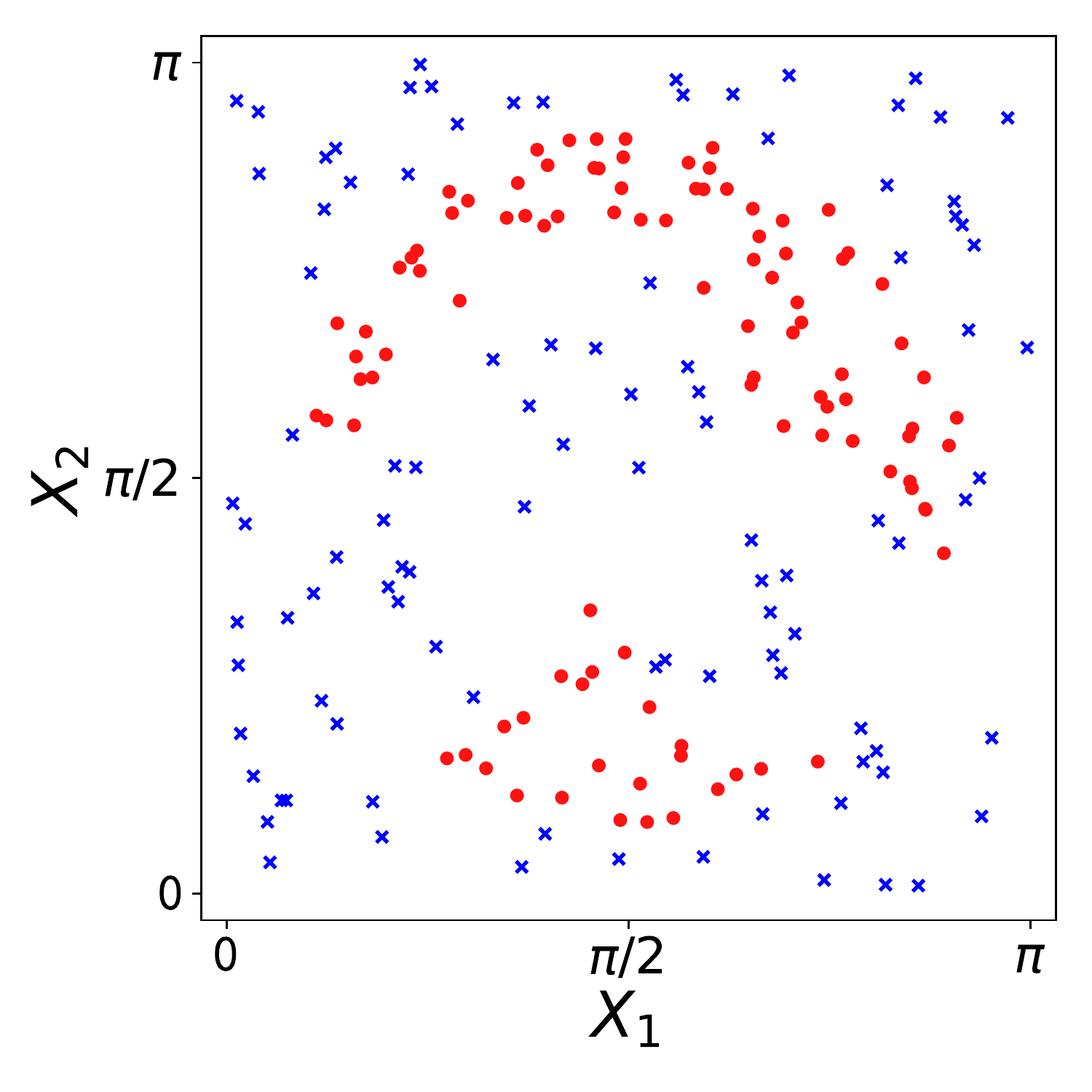}
}
\subfigure[]{
\label{subfig_c_complex}
\includegraphics[width=0.31\linewidth]{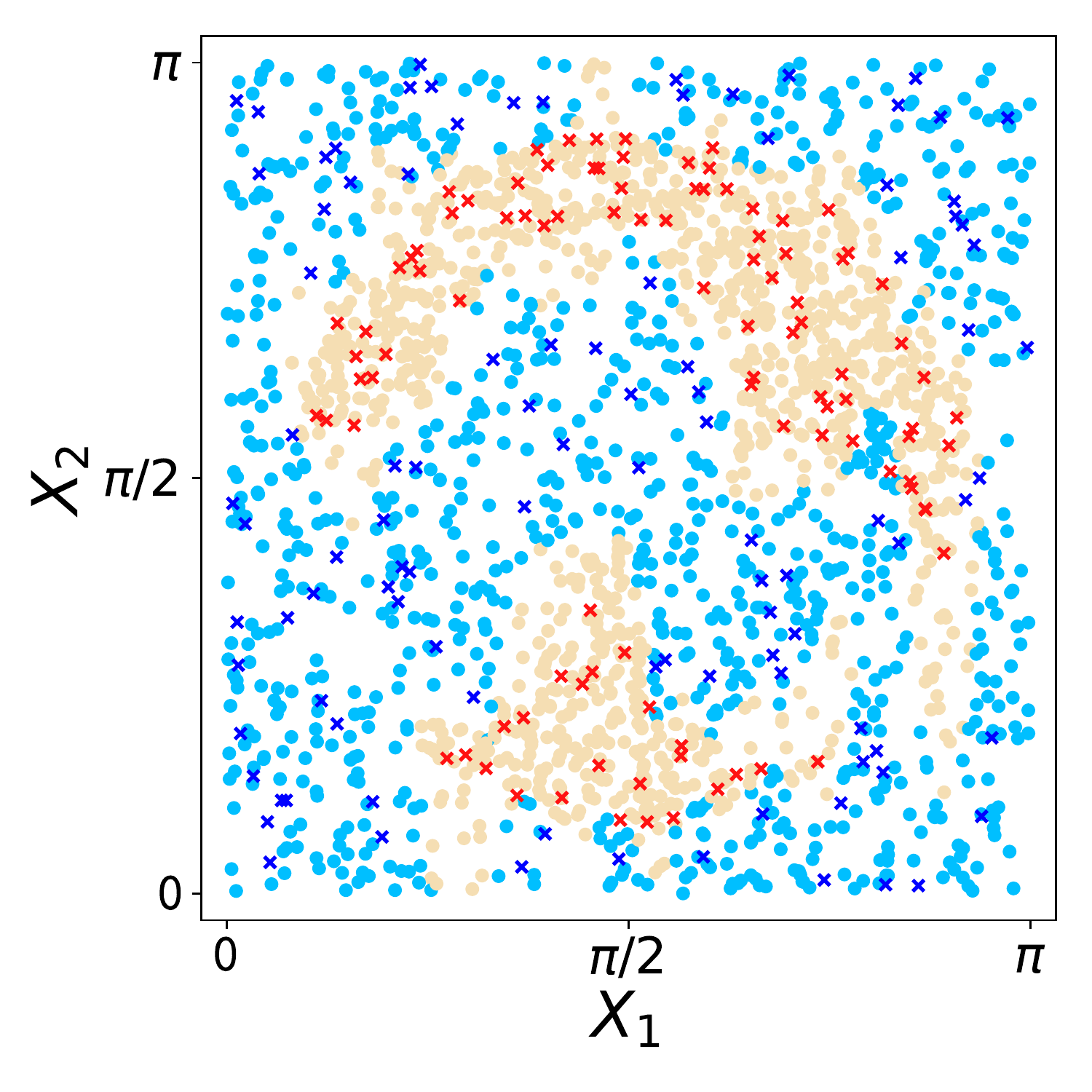}
}

\caption{{\bf Classification of randomly generated data:}
(a) Training data. In total, we generated 1100 blue points and 1100 red pints by the same generating function. 100 red points and 100 blue points are chosen randomly as training data, and the left pints are used as test data. (b) Training data including 100 red points are 100 blue points. (c) Prediction for the labels of test data. Light blue points are test data predicted as 'blue', and yellow points are predicted as 'red'. Red and blue cross represent training data.}
\label{fig_simu_complex}
\end{figure}

Here we will show another classification example, where the distribution of training data is artificially generated randomly.
Different from the example of $VO_2$, here the two groups can not be divided with a simple single boundary.
We generated 1100 red points and 1100 blue points at random, from which 100 red points and 100 blue points are picked up randomly as training data, the others will be used as test data.
    All test points are shown in Fig.(\ref{subfig_a_complex}), and the training points in Fig.(\ref{subfig_b_complex}).
    The two isolated groups of red points in  Fig.(\ref{subfig_a_complex}) are scattered along with the  blue ones covering  the whole area as shown in Fig.(\ref{subfig_b_complex}). The distribution of initial data makes it more challenging to classify the test data. After appropriate learning process, 54 sublabels (22 for red and 32 for blue) are obtained from the training data. Finally a classification circuit can be build with 7 qubits, 6 as control qubits and 1 for the given data. Prediction for labels of test data is shown in Fig.(\ref{subfig_c_complex}), where light blue dots are training points that are predicted as 'blue', yellow dots are predicted as 'red', red and blue cross represent the training data. Totally, 878 red test data and 836 blue test data are classified correctly, the matching rate for red and blue points can be estimated  as $87.8\%$ and $83.6\%$ respectively (when calculating match rate the training data are all excluded).

    \subsection{ Entanglement classification  in  Werner states}
    \label{Entanglement classification  in  Werner states}
    Further, our method can also be applied in entanglement classification.
    Consider the following scenario, Alice wants to send a  message to Bob, in which she will send  some entangled photon pairs of states  as digit 1 and some pairs at an untangled state as digit 0. Initially, Alice will send some photon pairs at various states for training by  informing   Bob which pair represents 1 and which represents 0.  Later she will use photon pairs states for communication. Although the widely used CHSH inequality\cite{clauser1969proposed} can be used to detect entanglement as the violation of CHSH inequality guarantees the existence of entanglement, however we can’t make any conclusion if the inequality is not violated. To address this issue consider Werner states in the density matrix form: 
    \begin{equation}
        \rho_W(p,\phi)=\frac{p}{4}|\Psi_B(\phi)\rangle\langle\Psi_B(\phi)|+\frac{1-p}{4}I
        \label{werner}
    \end{equation}
    where $I$ is  the $4\times4$  identity matrix, the parameter $0<p<1$, and   $|\Psi_B(\phi)\rangle$ is the Bell state given by:
    \begin{equation}
        |\Psi_B(\phi)\rangle=\frac{1}{\sqrt{2}}\left(
        |\uparrow\downarrow\rangle+e^{i\phi}|\downarrow\uparrow\rangle
        \right)
    \end{equation}
    
    Assume that Alice uses Werner states to transport information, while Bob will carry on the Bell test experiment with the following measurements: $Z, X; \frac{Z+X}{\sqrt{2}}$, and $ \frac{Z+X}{\sqrt{2}}$. From the measurement results, Bob will calculate four important correlation functions $E(Z, \frac{Z+X}{\sqrt{2}}), E(X, \frac{Z+X}{\sqrt{2}}), E(Z, \frac{Z-X}{\sqrt{2}}), E(X, \frac{Z-X}{\sqrt{2}})$ where
    \begin{equation}
        E(a,b)=\frac{N_{++}+N_{--}-N_{+-}-N_{-+}}{N_{++}+N_{--}+N_{+-}+N_{-+}}
    \end{equation}
    $N_{++}$ are the number of photon pairs whose measurement results are both $+1$ in the two channels. If Alice sets $\phi=0,\pi$, Bob will observe violation of CHSH inequality for $p>\frac{1}{\sqrt{2}}$. If Alice sets  $\phi=\pm\frac{\pi}{2}$, Bob can never observe violation of CHSH inequality. However, $\rho_W(p,\phi)$ will be entangled state when $p>\frac{1}{3}$. Consequently, CHSH inequality will not be a good classification way  for Bob. Instead, if Bob can set up a machine learning based on neural network, he will be able to 'decode' Alice's information with a much higher match rate\cite{gao2018experimental}. Here, we will show that our quantum classification algorithm can classify entanglement states  in such Werner state. We will take the 4-dimesinal vectors $E(Z, \frac{Z+X}{\sqrt{2}}), E(X, \frac{Z+X}{\sqrt{2}}), E(Z, \frac{Z-X}{\sqrt{2}})$, and  $E(X, \frac{Z-X}{\sqrt{2}})$ as input calculated  based on measurement results.
    
\begin{figure}[H]

\centering 

\subfigure[]{
\centering
\label{subfig_a_werner}
\includegraphics[width=0.45\linewidth]{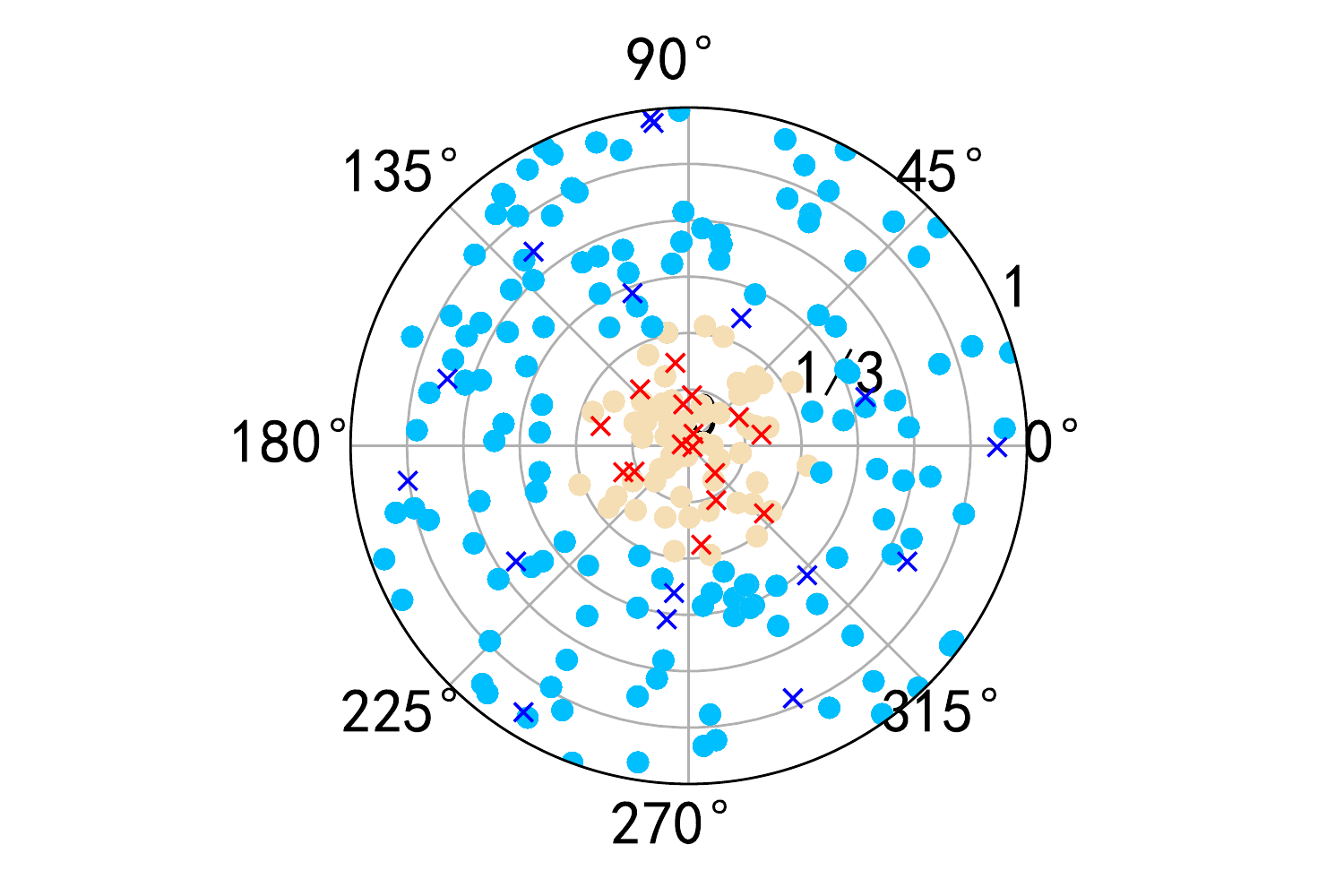}
}
\subfigure[]{
\centering
\label{subfig_b_werner}
\includegraphics[width=0.45\linewidth]{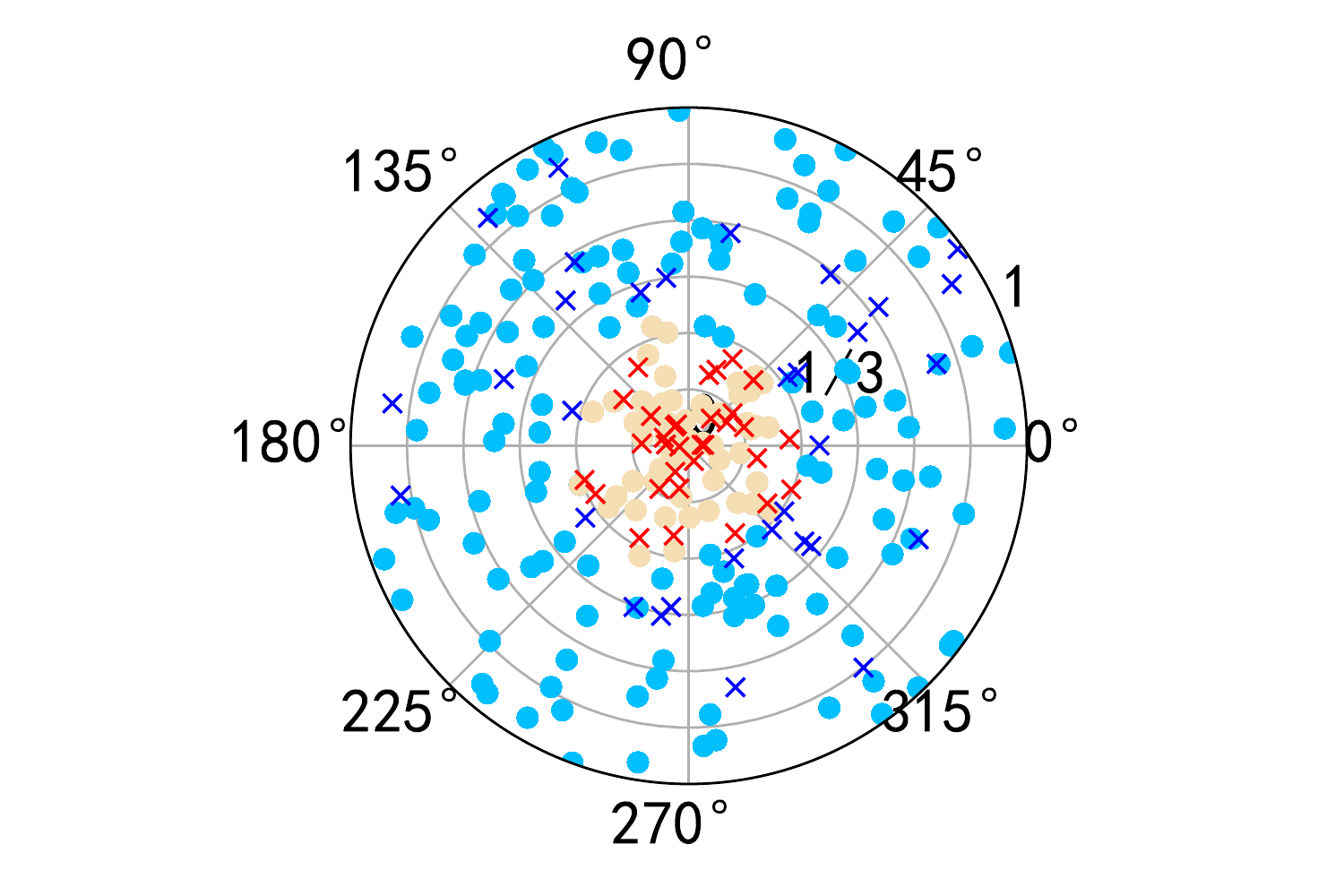}
}
\subfigure[]{
\label{subfig_c_werner}
\includegraphics[width=0.45\linewidth]{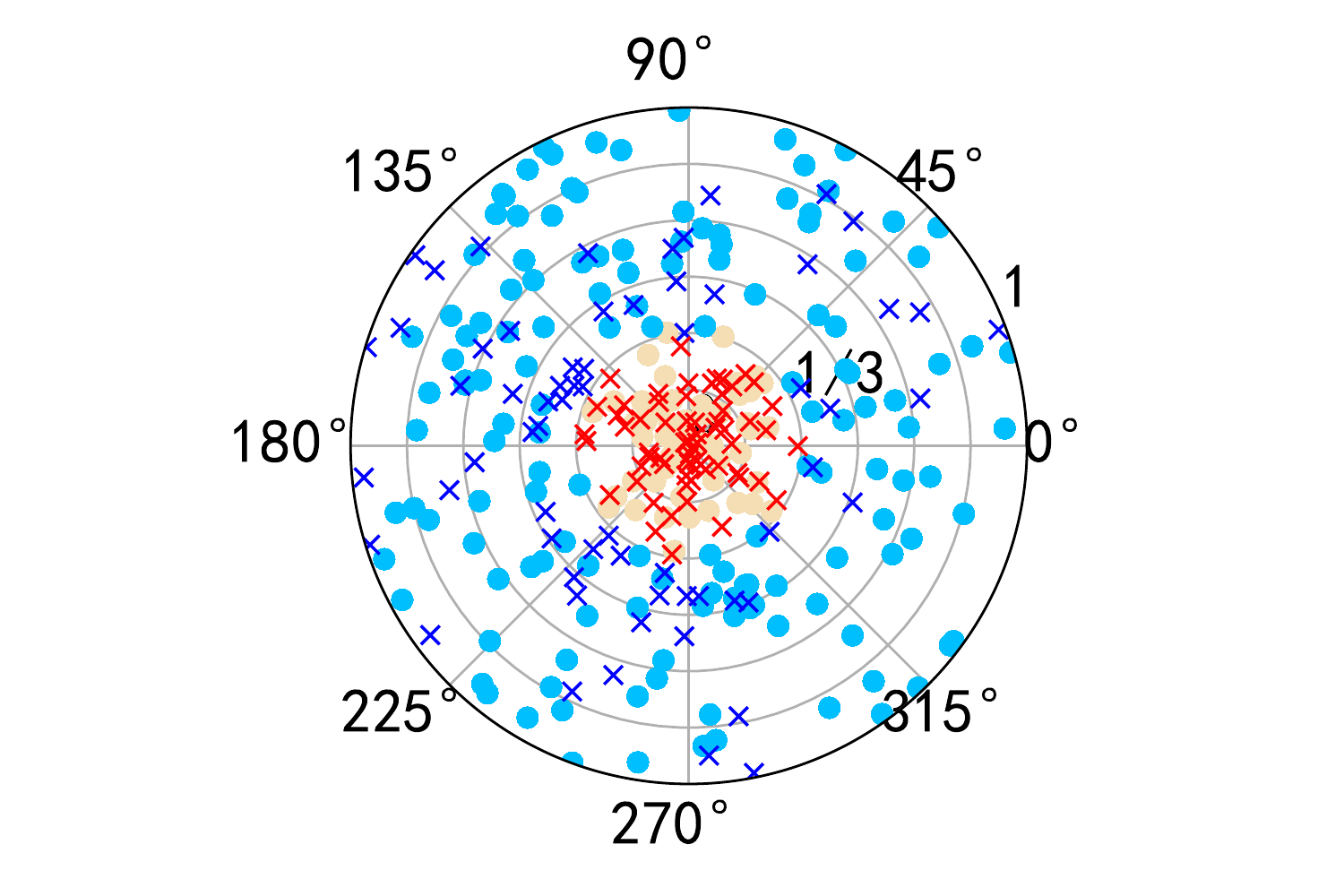}
}

\caption{{\bf Entanglement classification for Werner states:} In the plots, the radius represents the parameter $p$, and the angle represents  $\phi$. Every single dot represents a Werner state. Yellow dots represent test data that are predicted as 'untangled', and light blue dots represent test data predicted as 'entangled'. Cross represent training data, and red cross for untangled states, blue for entangled states. In all three figures we used the same 400 test data, half of which are entangled and the other half are untangled. Half of the training data are entangled states and the other half are untangled states. (a)32 test data are used, all are kept as sublabels. (b)64 test data are used, all are kept as sublabels. (c)128 test data are used, and only 64 of them are used as sublabels.}
\label{fig_simu_werner}
\end{figure}

    By changing $\phi$ and $p$ in Eq.(\ref{werner}), we can generate different entangled or untangled states. We prepared 200 entangled states and 200 untangled states as the test data. Moreover, we also generated a few different training data set, in each set half are entangled and the other half are untangled. After learning based on different training set, we can build quantum classification circuit to distinguish entanglement from the test data, and the simulation results are shown in Fig.(\ref{fig_simu_werner}). In (a), the training set only contains 32 points, and we keep them all as sublabels. So that there are $7$ qubits in the classification circuit (5 for sublabels and 2 for test data). 12 points are predicted with wrong label. In (b) the training set contains 64 points and all are kept as sublabels, and there are 8 qubits in the classification circuit. Finally 8 points are predicted with wrong labels. In (c) there are 128 points in the learning set and we derived 64 sublabels. Similarly as (b) we need 8 qubits to build the classification circuit, and only 6 qubits are predicted with wrong sublabels. However in (c) we do not plot the sublabels, as by applying our classification algorithms we can only get the parameters $E$ for every sublabel instead of $r, \phi$. 
    
    In these figures the points $(r,\phi)$ represent  Werner states.  Notice that $r, \phi$ are not used in the learning or classification process, as Bob does not know exact $r, \phi$ either. In the supplementary materials we  provide details of  the simulation.

    In the above discussion we assume that in communication between Alice and Bob, all measurement results are discrete, and one can easily calculate the parameters $E$. However, in chemical reactions the measurement results are often continuous, and one will get some special distributions after measurement. Recently, Zare and coworkers \cite{perreault2018cold} reported the rotationally inelastic collisions of deuterium hydride (HD, prepared at certain quantum states) with H$_2$ and D$_2$ under extremely low temperature (mean collision energy around 1 K), and they found that the orientation of HD molecules will lead to different distribution of scattering angle\cite{perreault2018cold}. 
    If the scattering experiment is applied as measurement to detect entanglement, then it would be nearly impossible to derive information about entanglement directly from the intricate raw data. Under such situations, some special methods, as we discussed in ref.\cite{li2019entanglement}, are required. With the assistance of auxiliary functions it will be possible to obtain the parameters $E$ from raw measurement results, after which by the same procedure we can build a quantum circuit for classification.

	\section{Further Discussions}
	\label{Further Discussions}
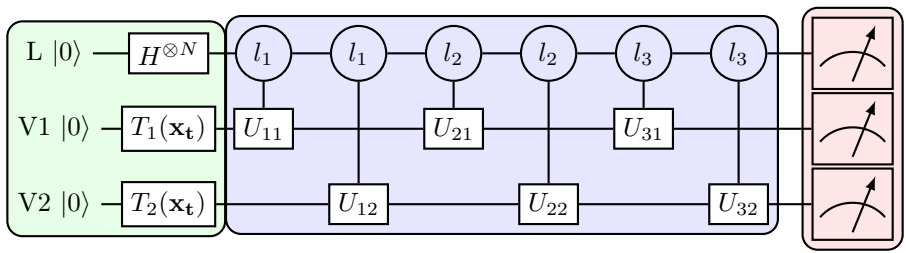
\begin{figure}[H]
			\centerline{
    \begin{tikzpicture}[thick]
    \tikzstyle{operator} = [draw,fill=white,minimum size=1.5em] 
    \tikzstyle{phase} = [fill,shape=circle,minimum size=5pt,inner sep=0pt]
    \tikzstyle{surround} = [fill=blue!10,thick,draw=black,rounded corners=2mm]
    \tikzstyle{sd_blue} = [fill=blue!10,thick,draw=black,rounded corners=2mm]
    \tikzstyle{sd_red} = [fill=red!10,thick,draw=black,rounded corners=2mm]
    \tikzstyle{sd_yellow} = [fill=yellow!10,thick,draw=black,rounded corners=2mm]
    \tikzstyle{sd_green} = [fill=green!10,thick,draw=black,rounded corners=2mm]
    %
    \node at (0,0) (q1){L\ \ket{0}};
    \node at (0,-1) (q2){V1\ \ket{0}};
    \node at (0,-2) (q3){V2\ \ket{0}};
    \node[operator] (op11) at (1.5,0) {$H^{\otimes N}$} edge [-] (q1);
    \node[operator] (op21) at (1.5,-1) {$T_1(\bf x_t)$} edge [-] (q2);
    \node[operator] (op31) at (1.5,-2) {$T_2(\bf x_t)$} edge [-] (q3);
    \node[circle0] (op12) at (2.75,0) {$l_1$} edge [-] (op11);
    \node[operator] (op22) at (2.75,-1) {$U_{11}$} edge [-] (op21);
    \draw[-] (op12) -- (op22);
    \node[circle0] (op13) at (4,0) {$l_1$} edge [-] (op12);
    \node[operator] (op33) at (4,-2) {$U_{12}$} edge [-] (op31);
    \draw[-] (op13) -- (op33);
    \node[circle0] (op14) at (5.25,0) {$l_2$} edge [-] (op13);
    \node[operator] (op24) at (5.25,-1) {$U_{21}$} edge [-] (op22);
    \draw[-] (op14) -- (op24);
    \node[circle0] (op15) at (6.5,0) {$l_2$} edge [-] (op14);
    \node[operator] (op35) at (6.5,-2) {$U_{22}$} edge [-] (op33);
    \draw[-] (op15) -- (op35);
    \node[circle0] (op16) at (7.75,0) {$l_3$} edge [-] (op15);
    \node[operator] (op26) at (7.75,-1) {$U_{31}$} edge [-] (op24);
    \draw[-] (op16) -- (op26);
    \node[circle0] (op17) at (9,0) {$l_3$} edge [-] (op16);
    \node[operator] (op37) at (9,-2) {$U_{32}$} edge [-] (op35);
    \draw[-] (op17) -- (op37);
    \node[meter] (meter1) at (10.5,0) {} edge [-] (op17);
    \node[meter] (meter2) at (10.5,-1) {} edge [-] (op26);
    \node[meter] (meter3) at (10.5,-2) {} edge [-] (op37);
    \begin{pgfonlayer}{background} 
    \node[sd_green] (background1) [fit = (q1) (op31)] {};
    \node[sd_blue] (background2) [fit = (op12) (op37)] {};
    \node[sd_red] (background3) [fit = (meter1) (meter3)] {};
    \end{pgfonlayer}
    \end{tikzpicture}
  }
	    \captionsetup{justification=centering}
	    \caption{
				{\bf Sketch of the main circuit for Method II:}	
				Qubits in the main circuit can be divided into two groups: One group will represent the sub labels, and will play the role of control bits, as the $L$ part in the figure. The other will represent the given vector, as the $V$ part in the circuit. Furthermore, qubits representing the vectors are divided into a few groups (In this Figure, 2 groups), and the sublabel qubits will control them respectively. We need to measure all of them to get our results. Connection between the $V$ qubits are not required in this circuit.}
		\label{device4}
		
	\end{figure}

	 So far, our classification has been restricted to two and four-dimensional vectors.Here, we discuss how to use it to classify vectors in higher dimensional space. Depending on the structure of qubits, two different mapping methods can be used: {\bf Mapping method I:} An arbitrary quantum state of N-qubits can be described as
	\begin{equation*}
	    |\Psi\rangle= \sum_{i=0}^{2^N-1}{c_i|i\rangle}
	\end{equation*}
	where $c_i$ can be rewritten as a function of ${\bf \Theta}=(\theta_1,\cdots,\theta_{2^N-1},\phi_1,\cdots,\phi_{2^N-1})$:
	\begin{align*}
	    &c_0 = \cos\theta_1\\
	    &c_1 = e^{i\phi_1}\sin{\theta_1}\cos{\theta_2}\\
	    &\cdots\\
	    &c_{2^N-2} = e^{i\phi_{2^N-2}}\Pi_{j=1}^{2^N-2}\sin{\theta_j}\cos{\theta_{2^N-2}}\\
	    &c_{2^N-1} = e^{i\phi_{2^N-1}}\Pi_{j=1}^{2^N-1}\sin{\theta_j}
	\end{align*}
	and $0\leq\theta_j\leq\pi$,$0\leq\phi_j\leq 2\pi$. Then there exists a  mapping operator $T(\bf \Theta)$, $|\Psi({\bf \Theta})\rangle=T(\bf \Theta)|0\rangle$, by which one could map a vector in $(2^N-1)$-dimensional  space into a quantum state of N-qubits. For this mapping method, the main structure of circuit is still like the one shown in Fig.(\ref{device2}). If all qubits in the main circuit are connected with others and we could build arbitrary quantum gates between any qubits in the main circuit, then we can obviously apply this mapping method I. However, sometimes connection in the machine could not satisfy our demand, then the second  mapping method might be more acceptable {\bf Mapping method II:} An arbitrary untangled quantum state of N-qubits can be described as:
	\begin{equation*}
	    \bigotimes_{j=1}^{N}{T_j({\bf x_t})|0\rangle} = \bigotimes_{j=1}^{N}{\left[\cos{\theta_j|0\rangle}+e^{i\phi_j}\sin{\theta_j}|1\rangle\right]}
	\end{equation*}
	Then we could map the vector ${\bf \Theta}=(\theta_1,\cdots,\theta_{N},\phi_1,\cdots,\phi_{N})$ into the untangled quantum state. Method II requires a circuit where the qubits representing sub labels are connected with all the qubits representing our vector, yet the connection between the 'data qubits' are not required. A sketch of the main circuit using method II can be found in fig.(\ref{device4}), where for simplicity, we note that $U_{ij}=U_j^{}(L_i) R_j(-\theta^{l_i}_{m}, -\phi^{l_i}_{m})$
	
	For the complexity analysis, let us assume that M d-dimensional vectors are offered as training data, thus $\lceil \log_2{d}\rceil$ qubits are required to represent the vectors. When measuring the inner product of two vectors, $\mathcal{O}\left\{\exp{(\log_2{d})}\right\}$ times of measurements are required. In the learning process to obtain sublabels, as discussed in sec.I, we need to repeat calculating inner products for $\mathcal{O}\left (M^2\right)$ times. Totally, the time complexity to obtain sublabels is $\mathcal{O}\left (M^2d\right)$. Then assume that we finally obtained  $L$ sublabels, then $\lceil \log_2{L}\rceil$ qubits are required to represent the sublabel. Thus, the quantum circuit to predict labels of test data contains $L$ multi-control gates. If we prepare the label qubits at uniform superposition state, then all qubits should be measured at last. For the label qubits, $\mathcal{O}\left\{\exp{(\log_2{L})}\right\}$ times of measurements are required, while for the data qubits $\mathcal{O}\left\{\exp{(\log_2{d})}\right\}$ times of measurements are required. As a result, we need to repeat measurement for $\mathcal{O}\left (Ld\right)$ times. However, if we prepare the label qubits at state eq.(\ref{lqubits}), then we only need to measure the first label qubit, and required times for measurement will be $\mathcal{O}\left (d\right)$.

	
	\section{Conclusion}
	\label{Conclusion}
	
	In summary, we developed a quantum classification algorithm where the training data is firstly clustered and assigned as various sublabels, and then based on these sublabels the quantum circuit is built for classification. 
	Further we applied this method for classifications of metallic-insulating  transition in $VO_2$, distinguish entanglement in Werner states, and classify some randomly generated data.
	Numerical simulation result shows that our algorithm is capable for various classification problems, especially the study of phases transition in materials.

\section*{Acknowledgement}
The authors would like to thank Dr. Masoud Mohseni, Dr. Manas Sajjan and Dr. Zixuan Hu for the helpful suggestions and discussions.
We acknowledge the financial support in part by the National Science Foundation under award number 1955907 and funding by the U.S. Department of Energy (Office of Basic Energy Sciences) under Award No.de-sc0019215.

	\bibliography{ref}

	\newpage
	\newcommand{\beginsupplement}{
	\setcounter{table}{0}
	\renewcommand{\thetable}{S\arabic{table}}
	\setcounter{figure}{0}
	\renewcommand{\thefigure}{S\arabic{figure}}
	\setcounter{equation}{0}
	\renewcommand{\theequation}{S\arabic{equation}}
}
	
	\beginsupplement
	\section*{Supplementary Materials}
	
	\subsection*{Details for the classification algorithm}
	The first process in the unsupervised learning process can be described by the following Algorithm.(\ref{algorithm1}) (Lloyd's algorithm\cite{mackay2003information}) :
	
	\begin{table}[H]
	    \centering
	    \begin{tabular}{r|l}
	    \hline\hline
	    \ & {\bf Algorithm S1: Basic training algorithm in the unsupervised learning process}\\
	    \hline
	    1 &{\bf INPUT} New data $\theta_{new},\phi_{new}$\\
	    2 &{\bf Existed Parameters: }The device $i$ represents data with same sub label $l_j$, $0\leq i\leq T$\\
	    3 &{\bf FOR} $0\leq i\leq T$ {\bf DO}\\
	    4 &\qquad $\theta_{1,i}=\theta_{m,i}$,\ $\phi_{1,i}=\phi_{m,i}$;\\
	    5 &\qquad $\theta_{2,i}=\theta_{new}$,\ $\phi_{2,i}=\phi_{new}$;\\
	    6 &\qquad {\bf MEASURE} Inner product $|\langle\psi(\theta_{1,i}, \phi_{1,i})|\psi(\theta_{2,i}, \phi_{2,i})\rangle$|\\
	    7 &{\bf END FOR}\\
	    8 &Find the maximum inner product $MAX\{ |\langle\psi(\theta_{1,j}, \phi_{1,j})|\psi(\theta_{2,j}, \phi_{2,j})\rangle$| \}\\
	    9 &{\bf IF} $MAX\{ |\langle\psi(\theta_{1,j}, \phi_{1,j})|\psi(\theta_{2,j}, \phi_{2,j})\rangle| \} < D$\\
	    10 &\qquad New data belongs to a new sub label. Set a new devise $T+1$ to represent this sub label.\\
	    11 &\qquad $\theta_{1,T+1}=\theta_{new}$,\ $\phi_{1,T+1}=\phi_{new}$;\ $N_{T+1}=1$\\
	    12 &{\bf ELSE} \quad $\theta_{1,j}=[\theta_{new}+N_j\theta_{m,j}]/(N_j+1)$,\ $\phi_{1,j}=[\phi_{new}+N_j\phi_{m,j}]/(N_j+1)$;\ $N_{j}=N_j+1$\\
	    \hline
	    \end{tabular}\\
	    \caption{Sketch of the basic training algorithm in the unsupervised learning process}
	    \label{algorithm1}
	\end{table}
	
	We need to input the training data to the prepared devices one by one, and for each new training data, Algorithm $S1$ is repeated for clustering the data. Note that there is an undefined parameter $D$, which is a constant given initially. $D$ is used to describe the minimum inner product between a state representing the average of a sub label and an arbitrary state representing training data with the same sub label, and by such a definition, 
	$ 0\leq D\leq 1 $. One might notice that if $D$ is too large, or in other word too close to 1, then algorithm $S1$ will lead to over-fitting, in which there are redundant sublabels. However, if $D$ is set too little, then our unsupervised learning process will be miserably useless, as we could not divide data with different patterns but the same prior label into suitable subgroups. In fact, the former is much more acceptable, so that initially, our suggestion is to choose the large $D$. To reduce the the redundancy, we could rely on an additional process as shown in Algorithm(\ref{algorithm2}).
	
	\begin{table}[H]
	    \centering
	    \begin{tabular}{r|l}
	    \hline\hline
	    \ & {\bf Algorithm S2: Basic algorithm to reduce redundancy}\\
	    \hline
	    1 &{\bf INPUT}  \ $\theta_i^{l_j},\phi_i^{l_j},N_i^{l_j}$, which represent average of the data with sub label i and prior label $l_j$\\
	    2 &{\bf New Parameters: }For each subgroup, set a new parameter $d_i^{l_j}=D$.\\
	    3 &{\bf FOR} Each subgroup ($\theta_i^{l_j},\phi_i^{l_j},N_i^{l_j}$) {\bf DO}\\
	    4 &\qquad\qquad {\bf FOR} All other sublabels with different prior label  ($\theta_k^{l_n},\phi_k^{l_n},N_k^{l_n}$, $n \neq j$) {\bf DO}\\
	    5 &\qquad\qquad\qquad $\theta_{1,i}=\theta_{m,i}$,\ $\phi_{1,i}=\phi_{m,i}$;
	    \ $\theta_{2,i}=\theta_{m,k}$,\ $\phi_{2,i}=\phi_{m,k}$;\\
	    6 &\qquad\qquad\qquad {\bf MEASURE} Inner product $|\langle\psi(\theta_{1,i}, \phi_{1,i})|\psi(\theta_{2,i}, \phi_{2,i})\rangle|$\\
	    7 &\qquad\qquad\qquad {\bf CALCULATE} $\Lambda =\cos\{\ \arccos{[\ |\langle\psi(\theta_{1,i}, \phi_{1,i})|\psi(\theta_{2,i}, \phi_{2,i})\rangle|\ ]}-\arccos{(d_k^{l_n})}\ \}$\\
	    8 &\qquad\qquad{\bf END FOR; CALCULATE $\Lambda_{max}$}\\
	    9 &\qquad\qquad {\bf FOR} All other sublabels with same prior label ($\theta_k^{l_j},\phi_k^{l_j},N_k^{l_j}$, $k \neq i$) {\bf DO}\\
	    10 &\qquad\qquad\qquad $\theta_{1,i}=\theta_{m,i}$,\ $\phi_{1,i}=\phi_{m,i}$;
	    \ $\theta_{2,i}=\theta_{m,k}$,\ $\phi_{2,i}=\phi_{m,k}$;\\
	    11 &\qquad\qquad\qquad {\bf MEASURE} Inner product $|\langle\psi(\theta_{1,i}, \phi_{1,i})|\psi(\theta_{2,i}, \phi_{2,i})\rangle$|\\
	    12 &\qquad\qquad\qquad {\bf CALCULATE} $\lambda =\cos\{\ \arccos{[\ |\langle\psi(\theta_{1,i}, \phi_{1,i})|\psi(\theta_{2,i}, \phi_{2,i})\rangle|\ ]}+\arccos{(d_k^{l_j})}\ \}$\\
	    13 &\qquad\qquad\qquad{\bf IF} $\lambda<\Lambda_{max}$\\
	    14 &\qquad\qquad\qquad\qquad $\theta_{1,i}=[N_i\theta_{m,i}+N_k\theta_{m,k}]/(N_i+N_k)$,\ $\phi_{1,i}=[N_i\phi_{m, i}+N_k\phi_{m,k}]/(N_i+N_k)$;\\
	    15 &\qquad\qquad\qquad\qquad$N_{i}=N_i+N_k;\ d_i^{l_j}=\cos\{\arccos(d_i^{l_j})+\lambda\}$\\
	    16 &\qquad\qquad\qquad\qquad{\bf RELEASE} $\theta_k^{l_j},\phi_k^{l_j},N_k^{l_j}$, {\bf BREAK FOR}\\
	    17 &\qquad\qquad{\bf END FOR}\\
	    18 &{\bf END FOR}\\
	    \hline
	    \end{tabular}\\
	    \caption{Sketch of the basic algorithm to reduce redundancy}
	    \label{algorithm2}
	\end{table}
	
	One might notice that in Algorithm (\ref{algorithm2}), we did not consider overlap between subgroups with different sublabels. Theoretically, one could always avoid such overlap by choosing small '$D$' initially. If we choose $D=1$ at start, then no overlap will be between these subgroups. However, to make our algorithm more efficient, $D$ should not be too small at the  beginning. Here we encourage to apply an additional algorithm to deal with subgroups with different sub labels,  as shown in the following Algorithm (\ref{algorithm3}), which shares some similarity  with Algorithm (\ref{algorithm2}). In fact, one could combine them together to reduce redundancy and overlap at the same time. The fundamental idea of Algorithm (\ref{algorithm3}) is to divide one subgroup that has overlap with others into a few smaller ones. Moreover, one could always rewrite line 15 as
	\begin{equation*}
	    {\bf IF} MAX\{|\langle\psi(\theta_{1,j}, \phi_{1,j})|\psi(\theta_{2,j}, \phi_{2,j})\rangle|\} < f(d_i^{l_j}), {\bf DO}
	\end{equation*}
	where $f(d_i^{l_j})$ is an arbitrary function ensuring that $d_i^{l_j}<f(d_i^{l_j})\leq 1$. A reminding is that if $d_i^{l_j}-f(d_i^{l_j})$ is too close to 0 then we might need to repeat Algorithm.(\ref{algorithm3}) many times to make sure that subgroups with different prior labels are untangled, while if $f(d_i^{l_j})$ itself being too close to 1 will lead to redundancy once again.

	\begin{table}[H]
	    \centering
	    \begin{tabular}{r|l}
	    \hline\hline
	    \ & {\bf Algorithm S3: Basic algorithm to reduce overlap}\\
	    \hline
	    1 &{\bf INPUT}  \ $\theta_i^{l_j},\phi_i^{l_j},N_i^{l_j}$, which represents average of the data with sub label i and prior label $l_j$\\
	    2 &{\bf New Parameters: }For each subgroup, set a new parameter $d_i^{l_j}=D$.\\
	    3 &{\bf FOR} Each subgroup ($\theta_i^{l_j},\phi_i^{l_j},N_i^{l_j}$) {\bf DO}\\
	    4 &\qquad\qquad {\bf FOR} All other sublabels with different prior label ($\theta_k^{l_n},\phi_k^{l_n},N_k^{l_n}$, $n \neq j$) {\bf DO}\\
	    5 &\qquad\qquad\qquad $\theta_{1,i}=\theta_{m,i}$,\ $\phi_{1,i}=\phi_{m,i}$;
	    \ $\theta_{2,i}=\theta_{m,k}$,\ $\phi_{2,i}=\phi_{m,k}$;\\
	    6 &\qquad\qquad\qquad {\bf MEASURE} Inner product $|\langle\psi(\theta_{1,i}, \phi_{1,i})|\psi(\theta_{2,i}, \phi_{2,i})\rangle|$\\
	    7 &\qquad\qquad\qquad {\bf CALCULATE} $\Lambda =\cos\{\ \arccos{[\ |\langle\psi(\theta_{1,i}, \phi_{1,i})|\psi(\theta_{2,i}, \phi_{2,i})\rangle|\ ]}-\arccos{(d_k^{l_n})}\ \}$\\
	    8 &\qquad\qquad\qquad {\bf IF} $\Lambda>d_i^{l_j}$, {Choose an arbitrary vector in this sublabel as average$\theta_{m,i},\phi_{m,i}$;}\\
	    9 &\qquad \quad \qquad\qquad{\bf FOR} other data with this sub label $\phi_{new}, \theta_{new}$, $0\leq i\leq T$,{\bf DO}\\
	    10 &\qquad\qquad\qquad\qquad $\theta_{1,i}=\theta_{m,i}$,\ $\phi_{1,i}=\phi_{m,i}$;\\
	    11 &\qquad\qquad\qquad\qquad $\theta_{2,i}=\theta_{new}$,\ $\phi_{2,i}=\phi_{new}$;\\
	    12 &\qquad\qquad\qquad\qquad {\bf MEASURE} Inner product $|\langle\psi(\theta_{1,i}, \phi_{1,i})|\psi(\theta_{2,i}, \phi_{2,i})\rangle$|\\
	    13 &\quad\qquad\qquad\qquad{\bf END FOR}\\
	    14 &\quad\qquad\qquad\qquad Find the maximum inner product $MAX\{|\langle\psi(\theta_{1,j}, \phi_{1,j})|\psi(\theta_{2,j}, \phi_{2,j})\rangle$|\}\\
	    15 &\quad\qquad\qquad\qquad{\bf IF} $MAX\{|\langle\psi(\theta_{1,j}, \phi_{1,j})|\psi(\theta_{2,j}, \phi_{2,j})\rangle|\} < \arccos( \frac{1}{2}d_i^{l_j})$, {\bf DO}\\
	    16 &\quad\qquad\qquad\qquad\qquad New data belongs to a new sub label. Set a new devise $T+1$.\\
	    17 &\quad\qquad\qquad\qquad\qquad $\theta_{1,T+1}=\theta_{new}$,\ $\phi_{1,T+1}=\phi_{new}$;\ $N_{T+1}=1$\\
	    18 &\quad\qquad\qquad\qquad{\bf ELSE}\\ 
	    19 &\quad\qquad\qquad\qquad\qquad$\theta_{1,j}=[\theta_{new}+N_j\theta_{m,j}]/(N_j+1)$,\ $\phi_{1,j}=[\phi_{new}+N_j\phi_{m,j}]/(N_j+1)$;\\
	    20 &\quad\qquad\qquad\qquad\qquad$N_{j}=N_j+1$\\
	    21 &\qquad\qquad{\bf END FOR}\\
	    22 &{\bf END FOR}\\
	    \hline
	    \end{tabular}\\
	    \caption{Sketch of the basic algorithm to reduce overlap between groups with different prior labels}
	    \label{algorithm3}
	\end{table}
	
	After repeating Algorithm (\ref{algorithm2},\ref{algorithm3}) several times, all the training vectors are divided into several sub groups with unique sublabels, and redundancy is deduced to minimum while sub groups with different prior labels are still distinguishable. Also,  we have  calculated their centroid vectors, which are stored in the devices as $\theta^{l_j}_{m}, \phi^{l_j}_{m}$. Next step is to build classification circuit based on these information, as discussed in the main article.
	
	\subsection*{Detailed example of  the quantum  classification algorithm}
\begin{figure}[H]

\centering 

\subfigure[]{
\centering
\includegraphics[width=0.35\linewidth]{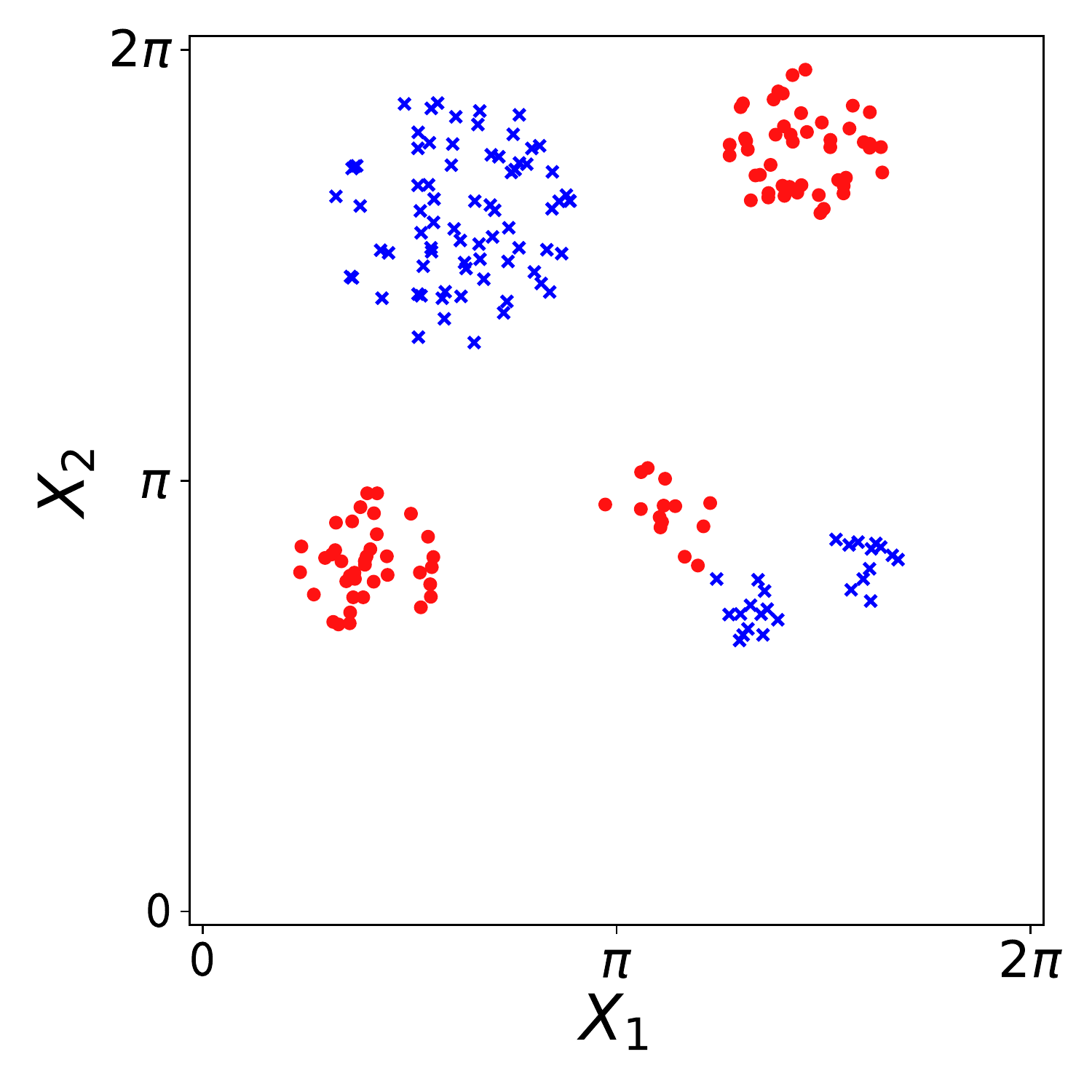}
}
\subfigure[]{
\centering
\includegraphics[width=0.35\linewidth]{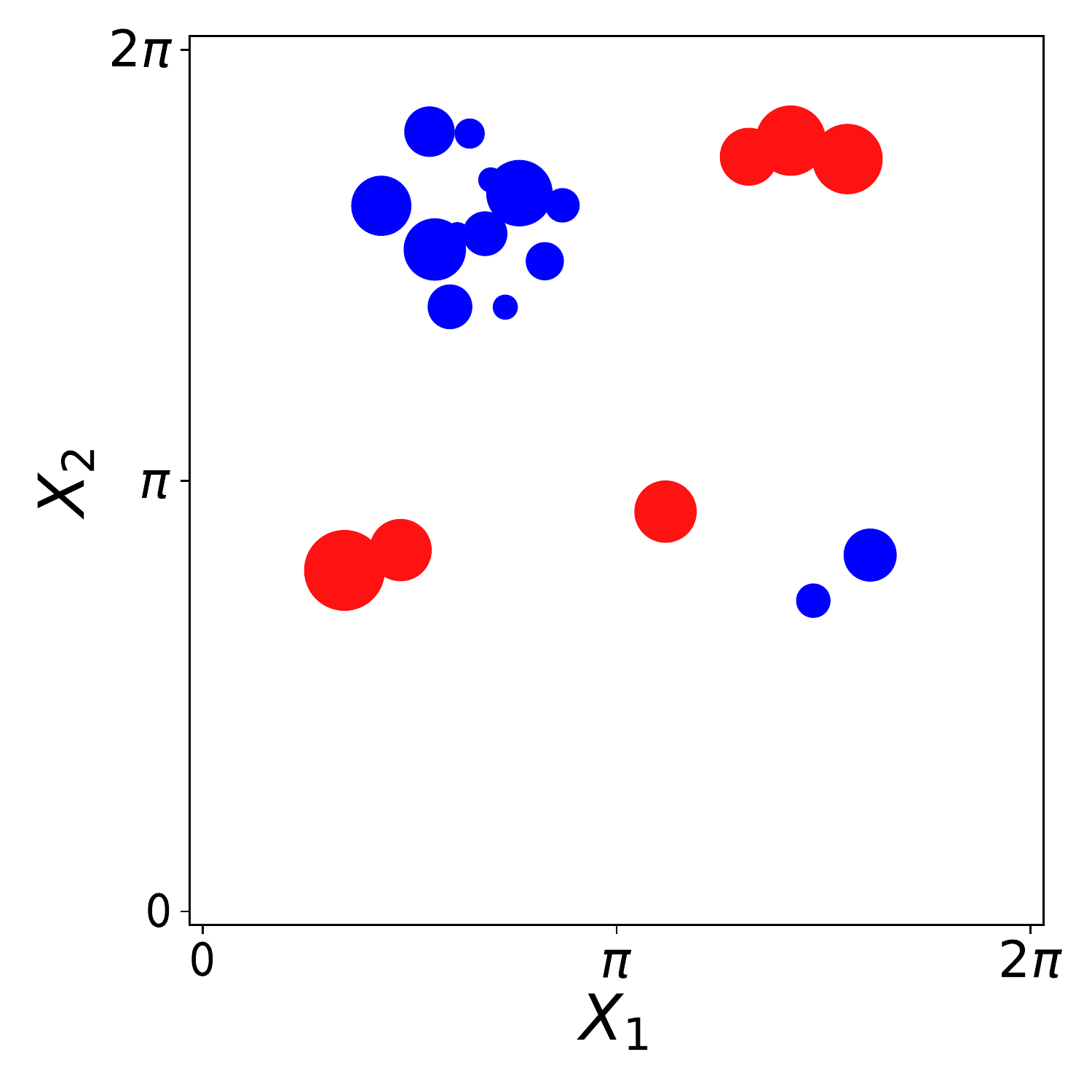}
}


\subfigure[]{
\label{subfig_c}
\includegraphics[width=0.35\linewidth]{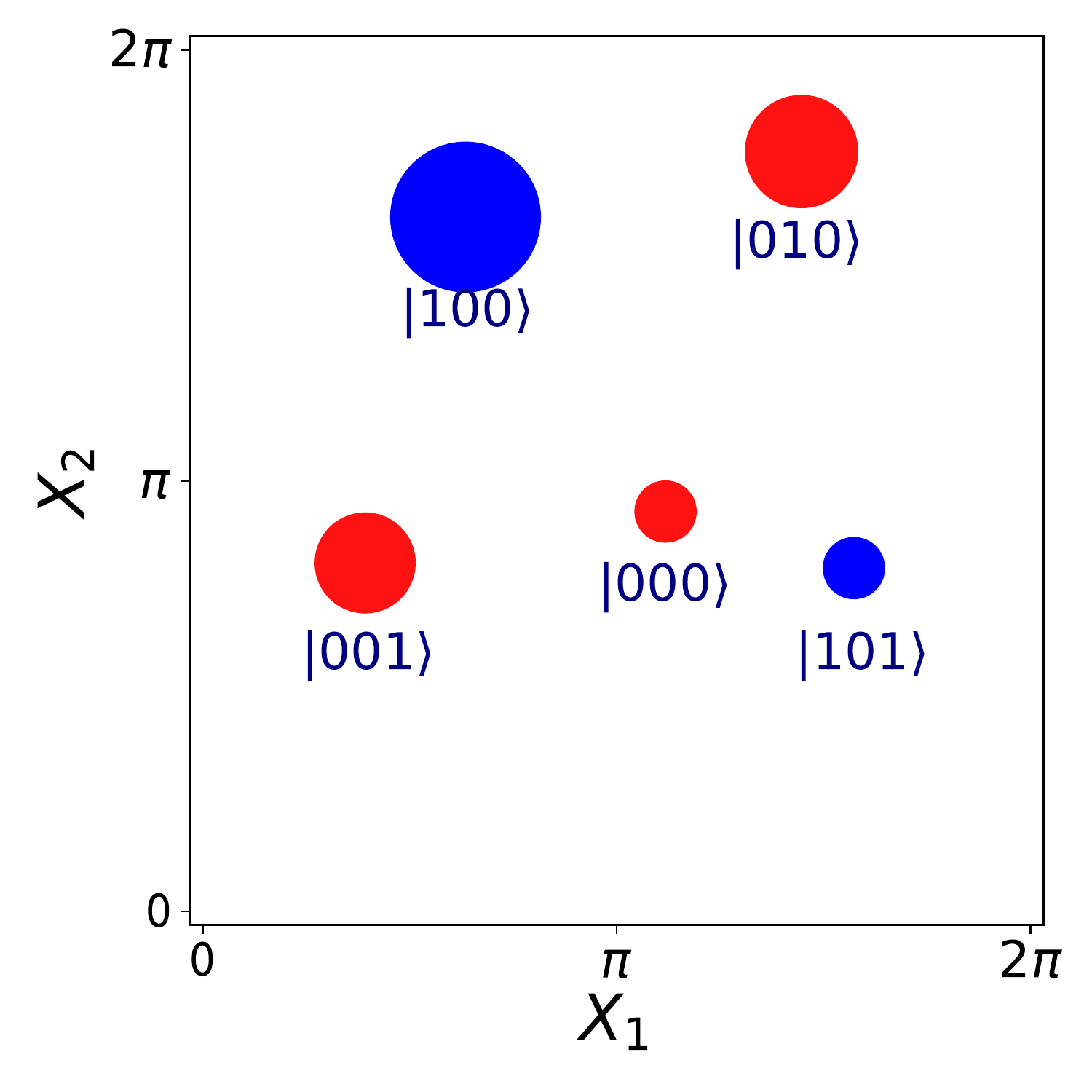}
}
\hspace{0.01\linewidth}
\subfigure[]{
\label{subfig_d}
\includegraphics[width=0.35\linewidth]{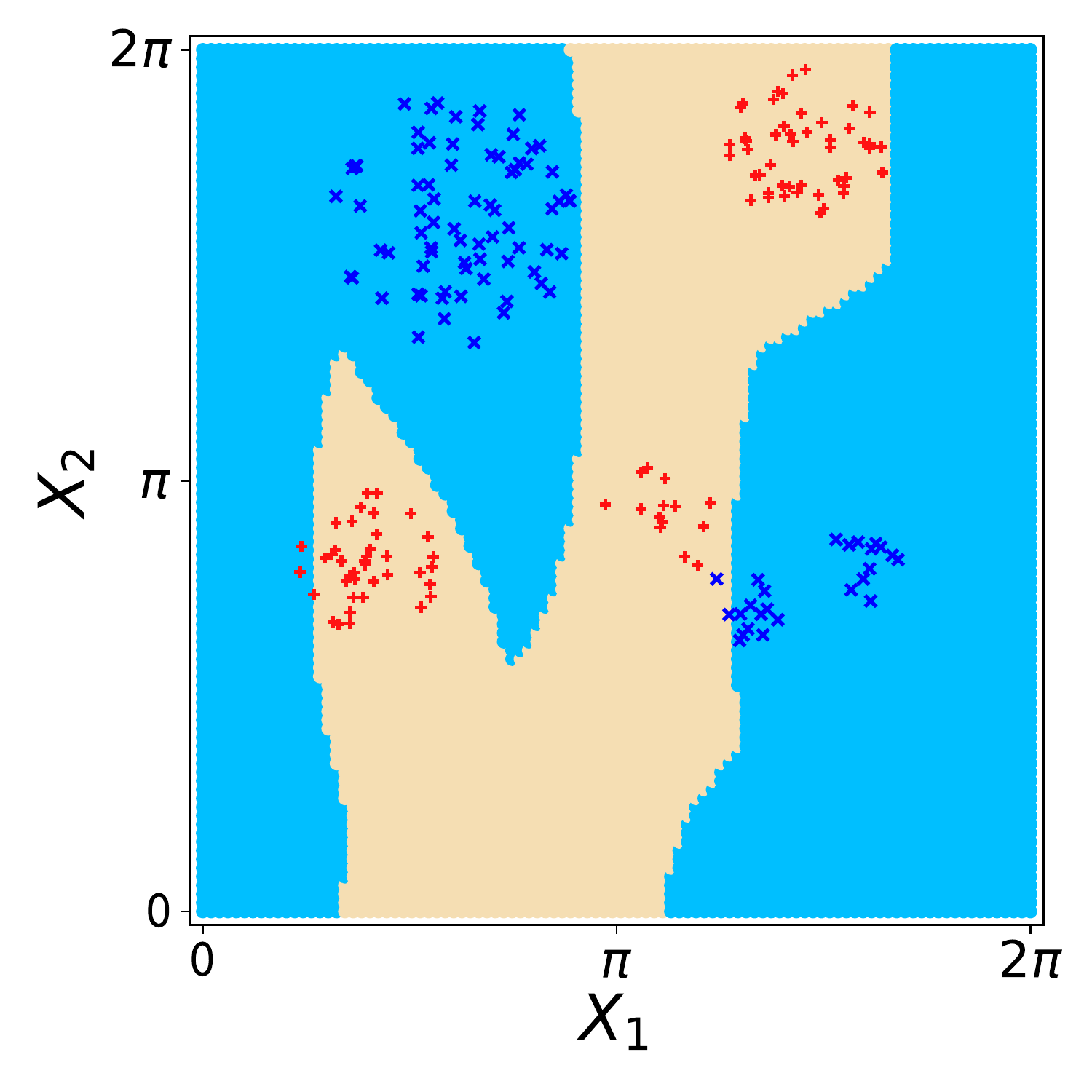}
}
\caption{{\bf Training data and simulation of the learning process:}
(a)The training vectors. Vectors in different color have different prior labels ('red' or 'blue'). (b)Subgroups we got after applying Algorithm.(\ref{algorithm1}). In fig.(\ref{fig_simu}b,c), we use blue or red sphere to represent data with the same sublabel, where the center of sphere represents the average vector, and the radius represents number of vectors belong to this sublabel. (c) Result after repeating Algorithm.(\ref{algorithm2},\ref{algorithm3}) a few times. (d)Prediction of new vectors. New vector in the blue part will be recognised with label 'blue', and label of new vectors in yellow part will be predicted as 'red'. }
\label{fig_simu}
\end{figure}

	In this section, we will offer an example to demonstrate how the Algorithms (\ref{algorithm1},\ref{algorithm2},\ref{algorithm3}) work. To get the training data, we generated some random vectors in 2-dimensional  space. These vectors are labeled as 'red' or 'blue', and color is prior label. Distribution of these data is shown in Fig.(\ref{fig_simu}a). One  note  that the red ones can be divided into a few subgroups and so can the blue ones. Firstly Algorithm (\ref{algorithm1}) is applied, as shown in shown in Fig.(\ref{fig_simu}b) the training data are divided into a great number of small subgroups. Redundancy appears because that we choose large $D=0.9$. To reduce excessive sublabels, next step is repeating Algorithm (\ref{algorithm2},\ref{algorithm3}) several times, after which we can get the results shown in Fig.(\ref{fig_simu}c). Now only five sublabels are left. In Fig.(\ref{fig_simu}b,c), we use blue or red spheres to represent data in different subgroups, where the center of sphere represents the average vector of a subgroup, and the radius of sphere represents number of vectors in the subgroup.

In Fig.[\ref{subfig_c}], we notice that we can not reduce the  number of sublabels further, and at  minimum 5 sublabels are required to describe the distribution of training data. As $2^2<5<2^3$, 3 qubits are required as "Label-qubit" which play roles as control qubit. In this case all training vectors are 2-d vectors, so that we need one qubit to represent a given vector. For at last the training vectors are divided into 5 subgroups, the main circuit will contain 5 “Control-Control-Control-U” (CCCU) gates. Based on the final left sublabels and their centroid vectors, classification circuit for new vectors can be built as Fig.(\ref{SM_fig_circuit}). 
\begin{figure}[H]
		\begin{center}
			\centerline{
    \begin{tikzpicture}[thick]
    \tikzstyle{operator} = [draw,fill=white,minimum size=1.5em] 
    \tikzstyle{phase} = [fill,shape=circle,minimum size=5pt,inner sep=0pt]
    \tikzstyle{phase0} = [draw, fill=white, shape=circle,minimum size=5pt,inner sep=0pt]
    \tikzstyle{cnot} = [draw, fill=white,shape=circle,minimum size=3pt,inner sep=0pt]
    \tikzstyle{surround} = [fill=blue!10,thick,draw=black,rounded corners=2mm]
    \tikzstyle{sd_blue} = [fill=blue!10,thick,draw=black,rounded corners=2mm]
    \tikzstyle{sd_red} = [fill=red!10,thick,draw=black,rounded corners=2mm]
    \tikzstyle{sd_yellow} = [fill=yellow!10,thick,draw=black,rounded corners=2mm]
    \tikzstyle{sd_green} = [fill=green!10,thick,draw=black,rounded corners=2mm]
    %
    \node at (-0.25,0) (q1) {q1\ \ket{0}};
    \node at (-0.25,-1) (q2) {q2\ \ket{0}};
    \node at (-0.25,-2) (q3) {q3\ \ket{0}};
    \node at (-0.25,-3) (q4) {q4\ \ket{0}};
    \node[operator] (op11) at (1,0) {$H$} edge [-] (q1);
    \node[operator] (op21) at (1,-1) {$H$} edge [-] (q2);
    \node[operator] (op31) at (1,-2) {$H$} edge [-] (q3);
    \node[operator] (op41) at (1,-3) {$T({\bf x_t})$} edge [-] (q4);
    \node[phase0] (op12) at (2.75,0) {} edge [-] (op11);
    \node[phase0] (op22) at (2.75,-1) {} edge [-] (op21);
    \draw[-] (op12)--(op22);
    \node[phase0] (op32) at (2.75,-2) {} edge [-] (op31);
    \draw[-] (op32)--(op22);
    \node[operator] (op42) at (2.75,-3) {$U_{000}$} edge[-] (op41);
    \node[phase0] (op13) at (4,0) {} edge [-] (op12);
    \node[phase0] (op23) at (4,-1) {} edge [-] (op22);
    \draw[-] (op13)--(op23);
    \node[phase] (op33) at (4,-2) {} edge [-] (op32);
    \draw[-] (op23)--(op33);
    \node[operator] (op43) at (4,-3) {$U_{001}$} edge[-] (op42);
    \draw[-] (op43)--(op33);
    \draw[-] (op42)--(op32);
    \node[phase0] (op14) at (5.25,0) {} edge [-] (op13);
    \node[phase] (op24) at (5.25,-1) {} edge [-] (op23);
    \draw[-] (op14)--(op24);
    \node[phase0] (op34) at (5.25,-2) {} edge [-] (op33);
    \draw[-] (op34)--(op24);
    \node[operator] (op44) at (5.25,-3) {$U_{010}$} edge[-] (op43);
    \draw[-] (op44)--(op34);
    \node[phase] (op15) at (6.5,0) {} edge [-] (op14);
    \node[phase0] (op25) at (6.5,-1) {} edge [-] (op24);
    \draw[-] (op15)--(op25);
    \node[phase0] (op35) at (6.5,-2) {} edge [-] (op34);
    \draw[-] (op25)--(op35);
    \node[operator] (op45) at (6.5,-3) {$U_{100}$} edge[-] (op44);
    \draw[-] (op45)--(op35);
    \node[phase] (op16) at (7.75,0) {} edge [-] (op15);
    \node[phase0] (op26) at (7.75,-1) {} edge [-] (op25);
    \draw[-] (op16)--(op26);
    \node[phase] (op36) at (7.75,-2) {} edge [-] (op35);
    \draw[-] (op26)--(op36);
    \node[operator] (op46) at (7.75,-3) {$U_{101}$} edge[-] (op45);
    \draw[-] (op46)--(op36);
    \node[meter] (meter1) at (9.25,0) {} edge [-] (op16);
    \node[meter] (meter2) at (9.25,-1) {} edge [-] (op26);
    \node[meter] (meter3) at (9.25,-2) {} edge [-] (op36);
    \node[meter] (meter4) at (9.25,-3) {} edge [-] (op46);
    \node[](archor) at (2.15, 0) {};
    \begin{pgfonlayer}{background} 
    \node[sd_green] (background1) [fit = (op11) (op41)] {};
    \node[sd_blue] (background2) [fit = (archor) (op46)] {};
    \node[sd_red] (background3) [fit = (meter1) (meter4)] {};
    \end{pgfonlayer}
    \end{tikzpicture}
  }
	\captionsetup{justification=centering}
	\caption{
			{\bf Sketch of the circuit for classification.}\\	
				$T({\bf x_t})$ converts state $|0\rangle$ into a quantum state representing the given test data ${\bf x_t}$. $q_{1,2,3}$ are label qubits, and $q_4$ represents the given vector. $U$ are single qubit control gates, and their parameters are derived in the previous steps.}
		\label{SM_fig_circuit}
		\end{center} 
	\end{figure}

 If one tries to simulate this main circuit on IBMQ 5-qubit machine, an auxiliary qubit is required. As not all qubits in the machine are connected with each other in IBM's machine. Assume that we can build arbitrary gates among $q_{1,2,3}$ or among $q_{3,4,5}$, yet $q_{1,2}$ and $q_{4,5}$ are not directly connected with each other, then fig.(\ref{device3}) can be a ketch for the 'CCCU' gates.

\begin{figure}[H]
		\begin{center}
			\centerline{
    \begin{tikzpicture}[thick]
    \tikzstyle{operator} = [draw,fill=white,minimum size=1.5em] 
    \tikzstyle{phase} = [fill,shape=circle,minimum size=5pt,inner sep=0pt]
    \tikzstyle{cnot} = [draw, fill=white,shape=circle,minimum size=3pt,inner sep=0pt]
    \tikzstyle{surround} = [fill=blue!10,thick,draw=black,rounded corners=2mm]
    %
    \node at (-0.25,0) (q1) {q1\ \ket{0}};
    \node at (-0.25,-1) (q2) {q2\ \ket{0}};
    \node at (-0.25,-2) (q3) {q3\ \ket{0}};
    \node at (-0.25,-3) (q4) {q4\ \ket{0}};
    \node at (-0.25,-4) (q5) {q5\ \ket{0}};
    \node[operator] (op11) at (1,0) {$H$} edge [-] (q1);
    \node[operator] (op21) at (1,-1) {$H$} edge [-] (q2);
    \node[operator] (op41) at (1,-3) {$H$} edge [-] (q4);
    \node[operator] (op51) at (1,-4) {$T({\bf x_t})$} edge [-] (q5);
    \node[operator] (op12) at (2,0) {$X$} edge [-] (op11);
    \node[operator] (op42) at (2,-3) {$X$} edge [-] (op41);
    \node[phase] (op13) at (2.75,0) {} edge [-] (op12);
    \node[phase] (op23) at (2.75,-1) {} edge [-] (op21);
    \draw[-] (op13)--(op23);
    \node[circlewc] (op33) at (2.75,-2) {} edge[-] (q3);
    \draw[-] (op23)--(op33);
    \node[phase] (op44) at (3.5,-3) {} edge[-] (op42);
    \node[operator] (op54) at (3.5,-4) {$V$} edge[-] (op51);
    \draw[-] (op44)--(op54);
    \node[phase] (op35) at (4.25, -2) {} edge[-] (op33);
    \node[circlewc] (op45) at (4.25, -3) {} edge[-] (op44);
    \draw[-] (op35)--(op45);
    \node[phase] (op46) at (5, -3) {} edge[-] (op44);
    \node[operator] (op56) at (5, -4) {$V^{\dagger}$} edge[-] (op54);
    \draw[-] (op46)--(op56);
    \node[phase] (op37) at (5.75, -2) {} edge[-] (op35);
    \node[circlewc] (op47) at (5.75, -3) {} edge[-] (op45);
    \draw[-] (op37)--(op47);
    \node[phase] (op38) at (6.5,-2) {} edge[-] (op37);
    \node[operator] (op58) at (6.5,-4) {$V$} edge[-] (op56);
    \draw[-] (op38)--(op58);
    \node[phase] (op19) at (7.25,0) {} edge [-] (op13);
    \node[phase] (op29) at (7.25,-1) {} edge [-] (op23);
    \draw[-] (op19)--(op29);
    \node[circlewc] (op39) at (7.25,-2) {} edge[-] (op38);
    \draw[-] (op29)--(op39);
    \node[operator] (op110) at (8,0) {X} edge [-] (op19);
    \node[operator] (op410) at (8,-3) {X} edge [-] (op47);
    \node(end1) at (9,0) {} edge [-] (op110);
    \node(end2) at (9,-1) {} edge [-] (op29);
    \node(end3) at (9,-2) {} edge [-] (op39);
    \node(end4) at (9,-3) {} edge [-] (op410);
    \node(end5) at (9,-4) {} edge [-] (op58);
    \begin{pgfonlayer}{background} 
    \node[surround] (b5ackground) [fit = (op12) (end5)] {};
    \end{pgfonlayer}
    \end{tikzpicture}
  }
	\captionsetup{justification=centering}
	\caption{
			{\bf Structure of the 'CCCU' gate in main circuit}\\	
				Here we show how to build the 'CCCU' gate (exactly, $CCCU_{010}$) in the main circuit with basic quantum gates, and the Toffoli gates can also be decomposed into a few CNOT gates and T gates (Or $\pi/8$ gate). As not all qubits in IBM Q machine are connected with each other, we introduce an auxiliary qubit (q3) to help us build the gate. q1, q2 and q4 are used to represent the sub label, and q5 is used to represent the given vector. $V$ is derived from the operator $U_{010}$, and 
				$U_{010}=V^2$.
				The $X$ gates are used as that we assume this sub label of is represented by state $|010\rangle$. $H$ gates are used as a preset of the sub label, and operator $T(\bf x_t)$ represents the mapping process.}
		\label{device3}
		\end{center} 
	\end{figure}
	
	Here we set all $|\Psi^f_L\rangle$ as $|0\rangle$. Then Parameters of gates $U$ satisfy that
	\begin{equation}
	    U_{i}|\Psi_{m,i}\rangle=|0\rangle
	\end{equation}
	where $|\Psi_{m,i}\rangle$ is state that represents the centroid state of all vectors with sublabel $i$. As an example, if a new vector ${\bf x_t}=(0.5\pi, 0.8\pi)$ is given as test vector, we can get results as following from IBM's simulator:
	\begin{center}
 \begin{tabular}{||c | c | c | c | c | c ||} 
 \hline
 $q_{1,2,4}$ &000 &001 &010 &100 &101 \\ [0.5ex] 
 \hline\hline
 $q_{5}=0$ &9.668\% & 10.84\% & 0.293\% & 3.125\% & 2.832\%\\ 
 \hline
\end{tabular}
\end{center}

	In our simulation, $000$ is sublabel for the central red subgroup, $001$ represents the red subgroup at left bottom, and $010$ represents the red subgroup at right upper right. $100$ represents the blue subgroup at upper left, and $101$ represents the blue subgroup at right bottom. So only the states shown in the table matters. For sublabel $q_1q_2q_4$, we can calculate $P(q_1q_2q_4;q_5=0)-P(q_1q_2q_4;q_5=1)$. The maximum $P(q_1q_2q_4;q_5=0)-P(q_1q_2q_4;q_5=1)$ corresponds to sublabel of this test vector. With results shown in the table, we know that the test vector is closer to subgroup $001$ than any other subgroups, obviously it is "red".
	
	Using the same method we can predict label for arbitrary vectors $[0, 2\pi]\times[0, 2\pi]$. Result is shown in fig.(\ref{fig_simu}d), where new vectors in light blue part will be recognised as label 'blue', and ones in the light yellow part will be predicted with label 'red'.
	
\subsection*{Classification study of the molecular HD scattering experiment}
    
\begin{figure}[H]

\centering 

\subfigure[]{
\centering
\label{subfig_a_scatter}
\includegraphics[width=0.3\linewidth]{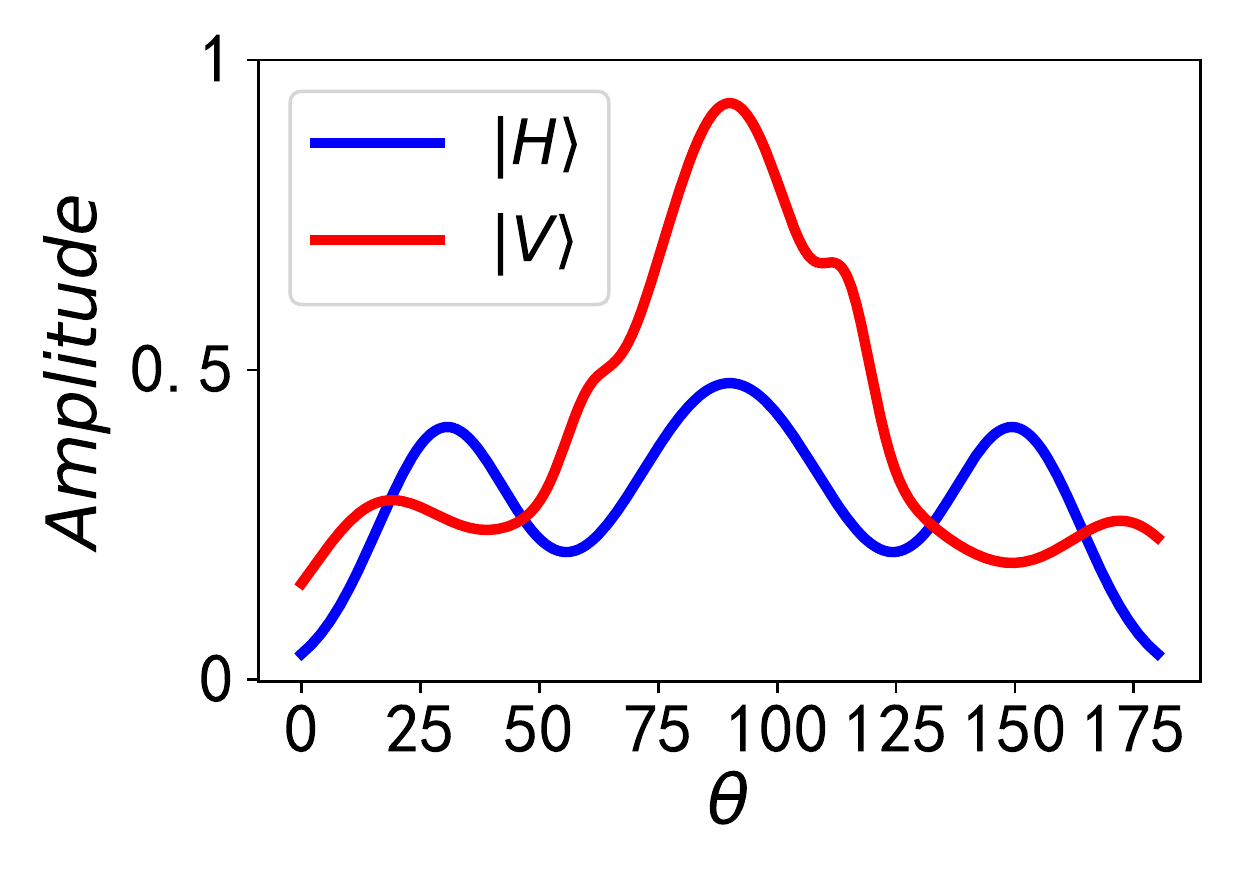}
}
\subfigure[]{
\centering
\label{subfig_b_scatter}
\includegraphics[width=0.3\linewidth]{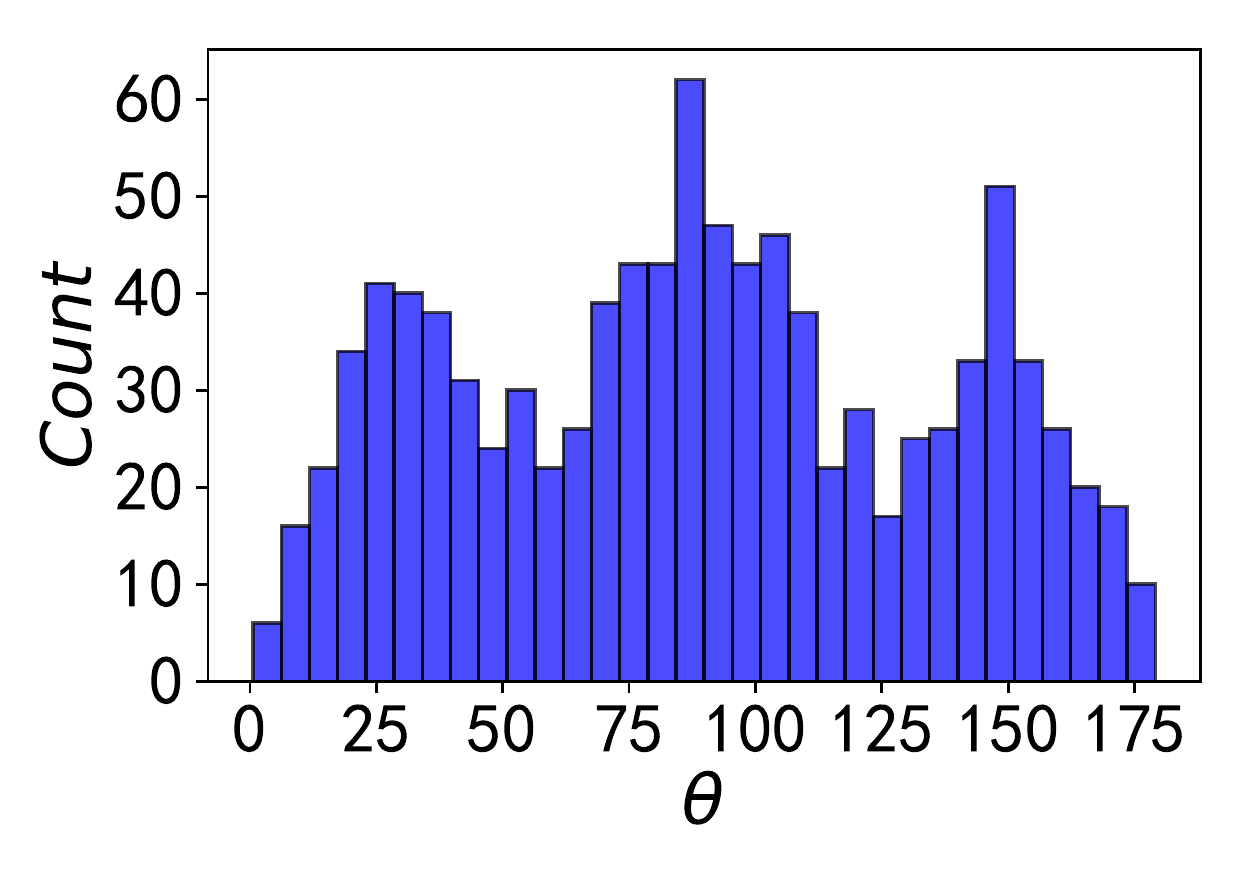}
}
\subfigure[]{
\label{subfig_c_scatter}
\includegraphics[width=0.3\linewidth]{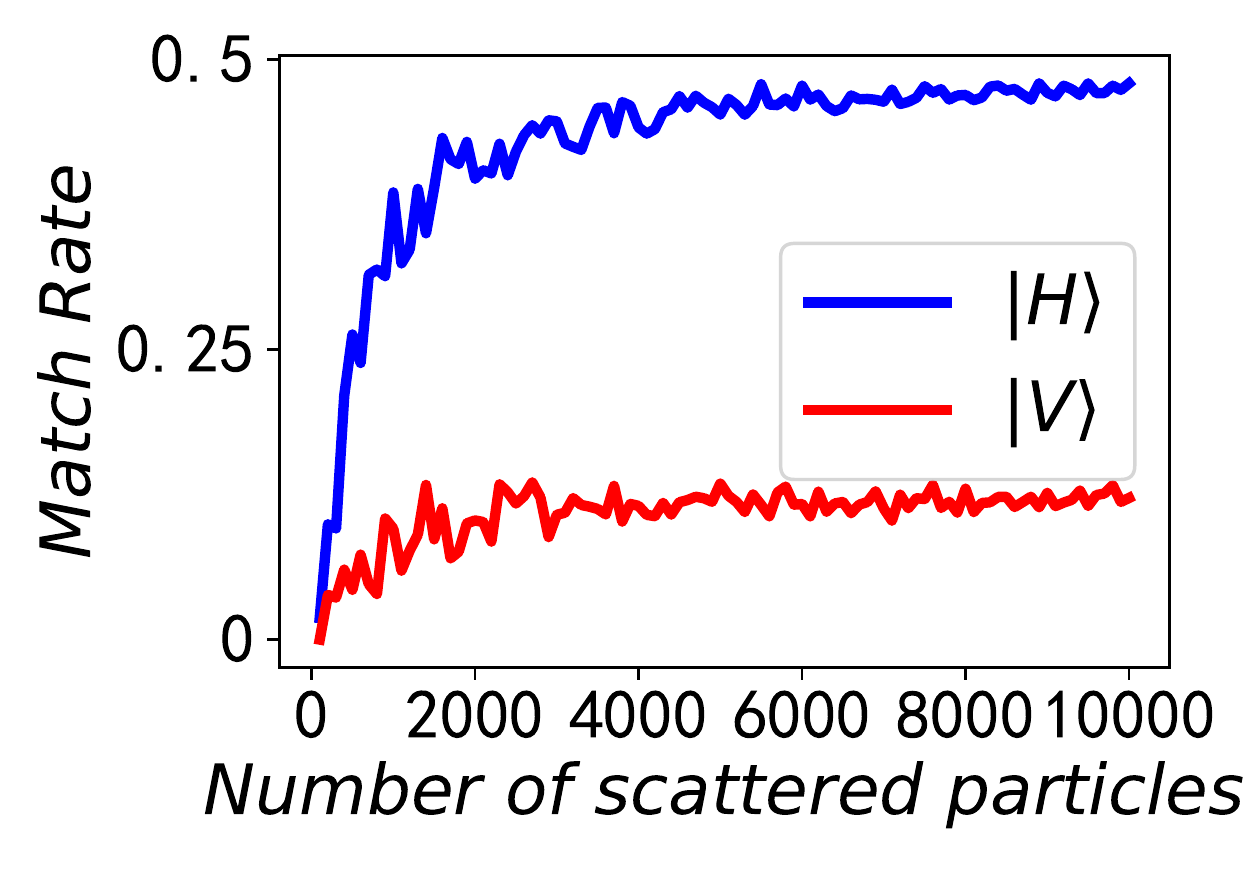}
}

\caption{{\bf Simulation results when studying scattering experiment:}
(a.) Standard distribution of scattering angle for initial state $|H\rangle$(blue) and $|V\rangle$(red). We plot this figure based on the experimental  results in ref.\cite{perreault2018cold}. (b) Distribution of scattering angle for initial state $|H\rangle$, only 1000 molecules are scattered. (c) Relationship between the matching  rate and number of scattered particles. All HD molecules are prepared initially at state $|H\rangle$.
}
\label{fig_simu_scatter}
\end{figure}

 In the recent scattering experiment of  the HD  molecule  with H$_2$ cluster,  states where the orientation of HD molecules is parallel to propagating direction $|H\rangle$  and states where the orientation of HD molecules is vertical to propagating direction $|V\rangle$  lead to two different scattering results.  As shown in fig.(\ref{subfig_a_scatter}) the blue and red curves represent the distribution of scattering angles  corresponding to the states $|H\rangle$  and $|V\rangle$  respectively \cite{perreault2018cold}. We can regard the initial state of the HD molecules as labels, and assume that we are provided with the standard results for state $|H\rangle$ and $|V\rangle$. Then if one carries on the same scattering experiment without knowing the initial state of HD, it is possible to use our quantum classification algorithm to distinguish the initial state of the HD molecule.  
    
    In our simulation, we used  6 qubits to build the quantum circuit for this classification, 1 qubit will represent the label and the other 5 qubits are used to describe the distribution of scattering angles, where the scattering angle is divided into $2^5=32$ slots. For instance, assume that there are 1000 HD molecules at the state $|H\rangle$ scattered with H$_2$ clusters, and distribution of these 1000 scattered particles is shown in fig.(\ref{subfig_b_scatter}). After counting particles in each slot, we could map the measurement result $f(\theta)$ as state $\psi[f(\theta)]$, and input state in the quantum circuit can be chosen as
    \begin{equation}
        \Psi[f(\theta)]=\frac{1}{\sqrt{2}}[|0\rangle+|1\rangle]\otimes\psi[f(\theta)].
    \end{equation}
    Finally the possibility to get measurement result as $|0\rangle\otimes|\psi^f_H\rangle$ is 0.389, while the possibility to get measurement result as $|1\rangle\otimes|\psi^f_V\rangle$ is 0.102, where we use state $|0\rangle$ and $|1\rangle$ of the label qubit to represent $|H\rangle$ and $|V\rangle$, and $|\psi^f_H\rangle$ and $|\psi^f_V\rangle$ are the final states for state $|H\rangle$ and $|V\rangle$ respectively. As 0.389 is significantly larger than 0.102, we can say that result shown in fig.(\ref{subfig_b_scatter}) infers to initial state $|H\rangle$.
    
    Furthermore, the  'Matching  rate' for $|H(V)\rangle$ is as the chance to get measurement result $|0(1)\rangle\otimes|\psi^f_{H(V)}\rangle$. For HD molecules at state $|H\rangle$, relationship between the matching rate and the number of scattered particles is shown in fig.(\ref{subfig_c_scatter}). When only a few particles are scattered, it is impossible to distinguish the initial state, and the matching  rate for both $|H\rangle$ and $|V\rangle$ are quite small. When more molecules are scattered, the pattern in the distribution will be clear, and the matching  rate for $|H\rangle$ will increase rapidly, until it is close to 0.5. Yet it can no more be greater than 0.5 because of fluctuation. On the other hand, the matching rate for $|V\rangle$ will increase at first and  finally be stable around 0.12, which infers the overlap between the standard distribution of state $|H\rangle$ and $|V\rangle$.

	\subsection*{Simulation results for phase transition in VO$_2$}
\begin{figure}[H]

\centering 

\includegraphics[width=0.7\linewidth]{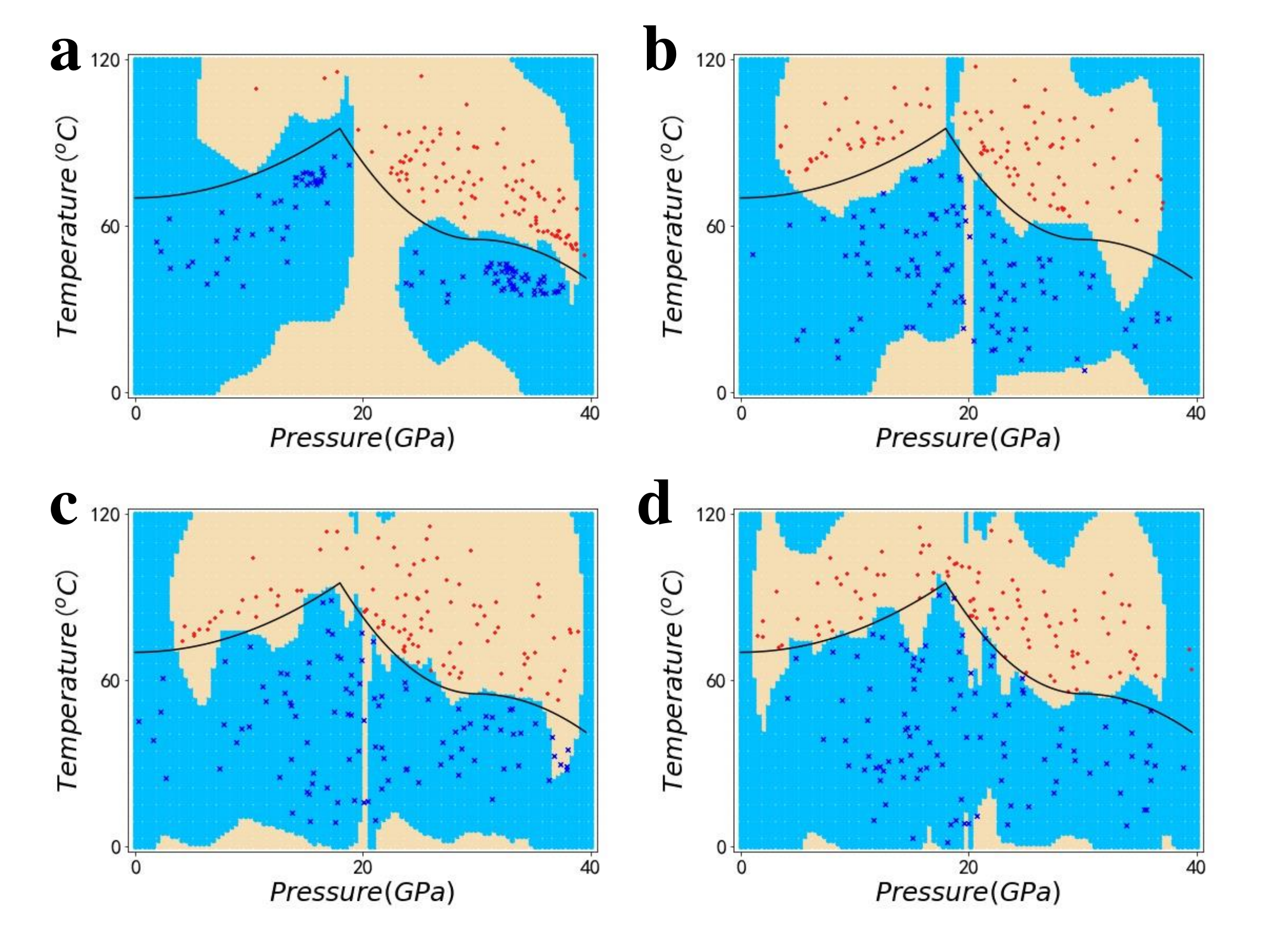}

\caption{{\bf Simulation results for classification of  phase transition in VO$_2$:} Dots in different colors are used as training data. Red ones are at metallic  states, and blue ones are at insulating states.
(a)Instead of choosing training data randomly, here we picked training vectors artificially. For metallic states, more data at the right side are used and only few at the left side. For insulating state, we avoid using data around $P=20 GPa$.
(b,c)Training data are randomly picked in a larger scale (More vectors are very far away from the black curve). In (c), we use data a little closer to the black curve comparing to (a) and (b). 
(d)Training data are randomly picked in a larger scale, and now we also use data very close to the black curve, some points are just on the curve. }
\label{SM_fig_multi}
\end{figure}

	Here we would like to offer more simulation results when studying the phase transition, as shown in fig.(\ref{SM_fig_multi}).
	
	We can notice that for all these four training data sets, our algorithm can still classify different states efficiently. In fig.(\ref{SM_fig_multi}a) training vectors are chosen artificially instead of randomly, and we can notice that the predicted boundary is no more close to black curve. When we starts choosing data randomly, as in fig.(\ref{SM_fig_multi}b,c,d), the predicted boundary between metallic  and insulating states will still be around the black curve. Further, when we use data closer to the black curve, the prediction will be more accurate, as the results shown in fig.(\ref{SM_fig_multi}d) is better than in fig.(\ref{SM_fig_multi}b,c).
	
	\subsection*{Simulation results for randomly generated data}
	
\begin{figure}[H]

\centering 

\subfigure[]{
\centering
\includegraphics[width=0.35\linewidth]{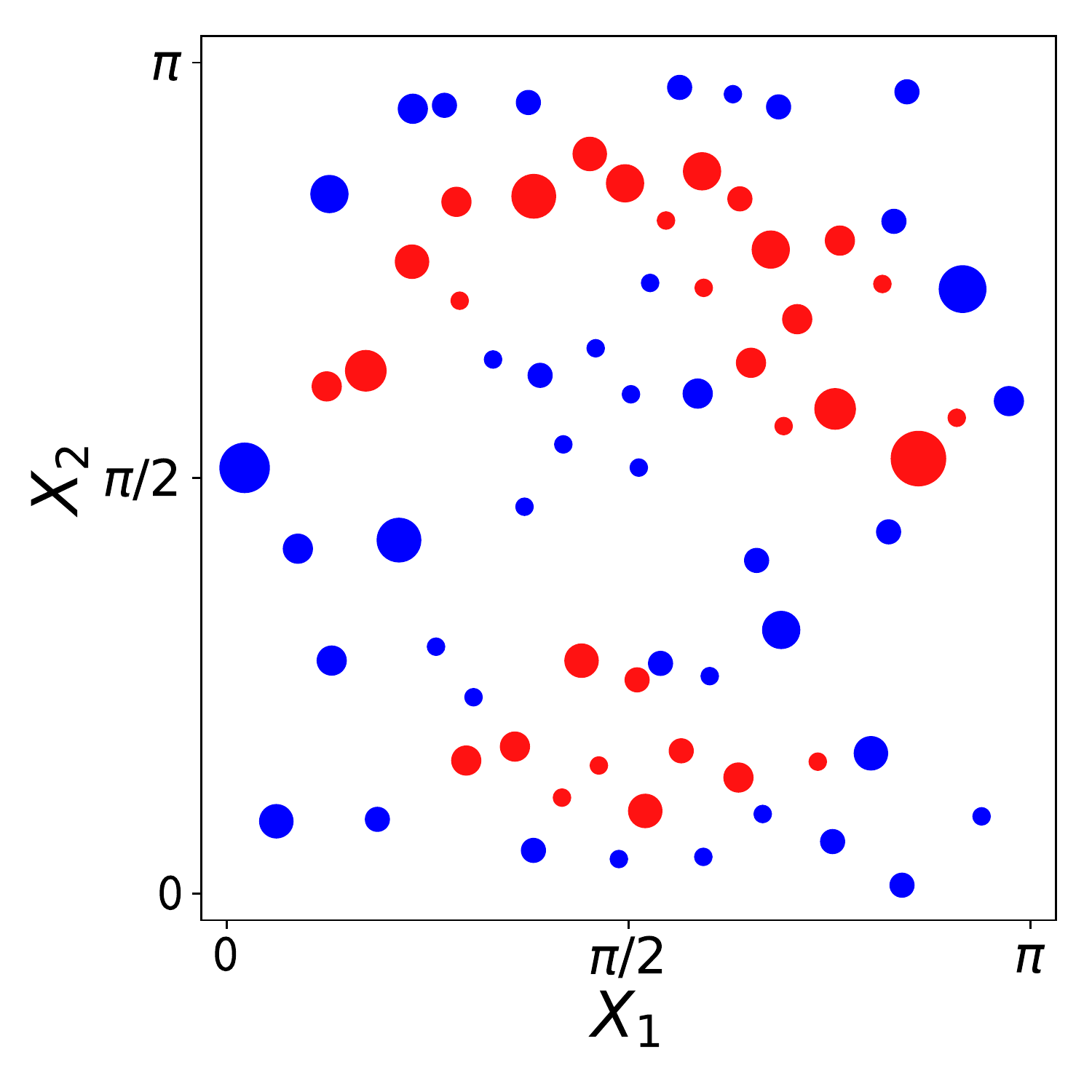}
}
\subfigure[]{
\centering
\includegraphics[width=0.35\linewidth]{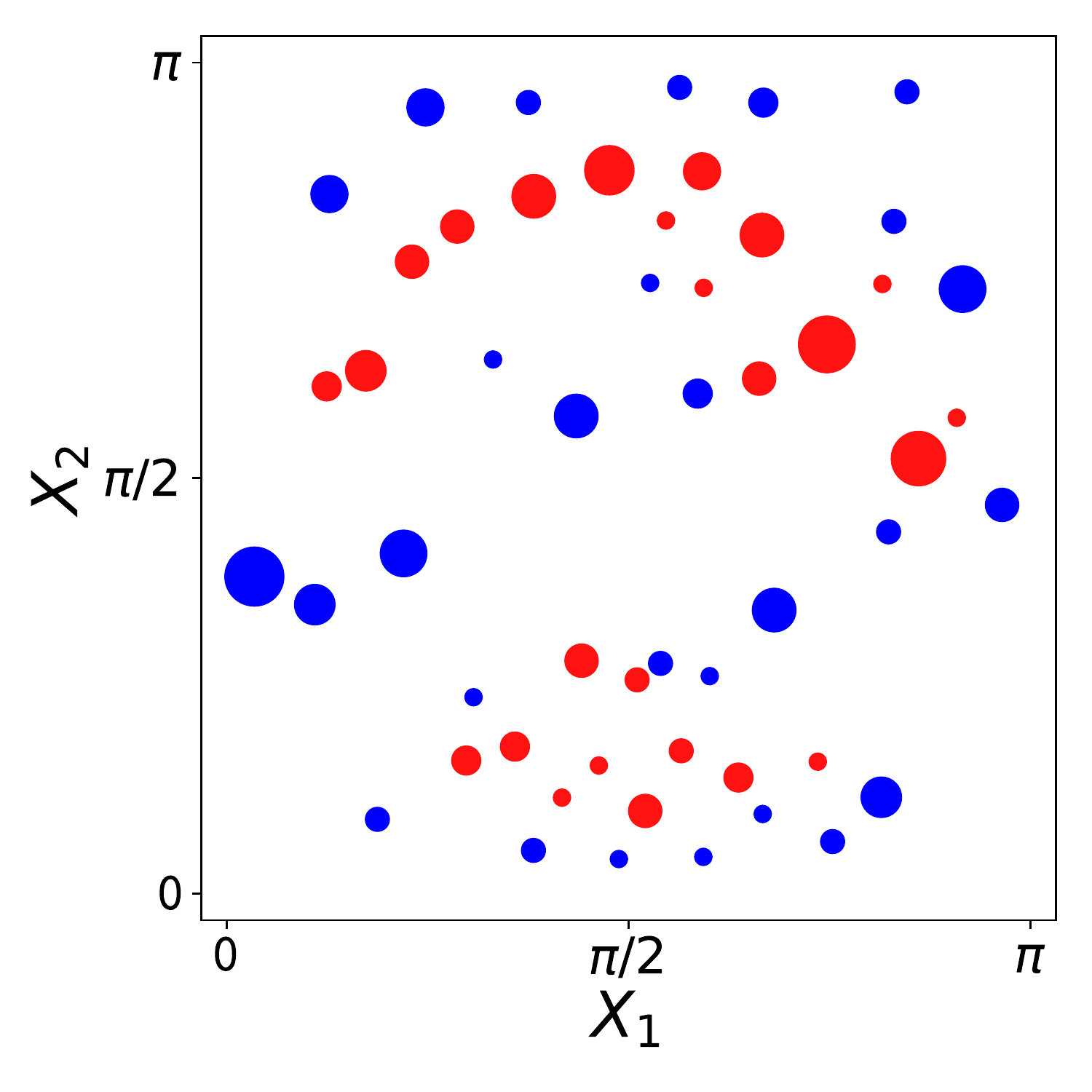}
}

\caption{{\bf Intermediate simulation results when classifying artificially generated data:} (a) Sublabels left after applying the clustering algorithm. (b) Sublabels left after repeating adjusting algorithms a few times. There are 22 red sublabels and 32 blue sublabels. Similarly, we use blue or red sphere to represent data with the same sublabel, where the center of sphere represents the average, and the radius represents number of vectors belong to this sublabel.}
\label{sm_simu_complex}
\end{figure}

	Intermediate simulation results are shown in fig.(\ref{sm_simu_complex}). (a) shows results after applying the cluster algorithm, and (b) shows sublabels left after repeating the adjust algorithms to reduce redundancy. As we mentioned in the main text, finally 22 red sublabels and 32 blue sublabels are left. Though we can reduce numbers of sublabels further, we can not reduce the number of total sublabels to 32 or less without losing important information. Thus,  we decided to keep these 54 sublabels and build classification algorithm based on them.

	\subsection*{ Entanglement classification in Werner states}

	Consider the measurement sets $Z, X; \frac{Z+X}{\sqrt{2}}, \frac{Z+X}{\sqrt{2}}$, for Bell state $|\Psi_+\rangle=|\Psi_B(\phi=0)\rangle$,
	\begin{equation}
        |\Psi_+\rangle=\frac{1}{\sqrt{2}}\left(
        |\uparrow\downarrow\rangle+|\downarrow\uparrow\rangle
        \right)
    \end{equation}
	Theoretically, the expectation value of $E$ are
	\begin{equation*}
	    \begin{split}
	        E(Z, \frac{Z+X}{\sqrt{2}})=\frac{1}{\sqrt{2}}\\
	        E(Z, \frac{Z-X}{\sqrt{2}})=\frac{1}{\sqrt{2}}\\
	        E(X, \frac{Z+X}{\sqrt{2}})=-\frac{1}{\sqrt{2}}\\
	        E(X, \frac{Z-X}{\sqrt{2}})=\frac{1}{\sqrt{2}}
	    \end{split}
	\end{equation*}
	and 
	\begin{equation}
	    E(\phi=0)=\left|E(Z, \frac{Z+X}{\sqrt{2}})+
	        E(Z, \frac{Z-X}{\sqrt{2}})-
	        E(X, \frac{Z+X}{\sqrt{2}})+
	        E(X, \frac{Z-X}{\sqrt{2}})\right|=2\sqrt{2}
	\end{equation}
	reaches a maximal and CHSH inequality is violated. However, for $|\Psi_B(\phi=\frac{\pi}{2})\rangle$,
	\begin{equation}
        |\Psi_B(\phi=\frac{\pi}{2})\rangle=\frac{1}{\sqrt{2}}\left(
        |\uparrow\downarrow\rangle+i|\downarrow\uparrow\rangle
        \right)
    \end{equation}
    Though there is still the maximum entanglement, one can never observe violation of CHSH inequality theoretically, as now
    \begin{equation*}
	    \begin{split}
	        E(Z, \frac{Z+X}{\sqrt{2}})=\frac{1}{\sqrt{2}}\\
	        E(Z, \frac{Z-X}{\sqrt{2}})=\frac{1}{\sqrt{2}}\\
	        E(X, \frac{Z+X}{\sqrt{2}})=0\\
	        E(X, \frac{Z-X}{\sqrt{2}})=0
	    \end{split}
	\end{equation*}
	and 
	\begin{equation}
	    E(\phi=0)=\left|E(Z, \frac{Z+X}{\sqrt{2}})+
	        E(Z, \frac{Z-X}{\sqrt{2}})-
	        E(X, \frac{Z+X}{\sqrt{2}})+
	        E(X, \frac{Z-X}{\sqrt{2}})\right|=\sqrt{2}
	\end{equation}
    
    Consequently, though it is a universal method, for certain situation CHSH inequality might not be the best choice to distinguish entanglement. 
    
    {\bf Mapping strategy in the simulation:} Different from the former examples, here the input (test vectors or learning vectors) will have 4 components, as a result 2 qubits are required to represent a vector (denoted  $q_1$ and $q_2$). As we discussed in the main text, there are two different mapping methods and the later one is chosen. The classification circuit has the same structure with the one shown in Fig.(6c). In the $CU$ gates the $U$ will always be single qubit rotation gates. Firstly, we apply mapping method:
    \begin{equation*}
        \begin{split}
             E(Z, \frac{Z+X}{\sqrt{2}})\rightarrow \theta_1\\
	        E(Z, \frac{Z-X}{\sqrt{2}})\rightarrow \phi_1\\
	        E(X, \frac{Z+X}{\sqrt{2}})\rightarrow \theta_2\\
	        E(X, \frac{Z-X}{\sqrt{2}})\rightarrow \phi_2
        \end{split}
    \end{equation*}
	
	Then go through the learning process and build classification circuit. For every test data, collect the measurement results, and denote $P_1(l_i)$ as the possibility to find qubits at aim state corresponding to sublabel $l_i$.
	
	Next, we will apply another mapping method given by:
	\begin{equation*}
        \begin{split}
             E(Z, \frac{Z+X}{\sqrt{2}})\rightarrow \phi_1\\
	        E(Z, \frac{Z-X}{\sqrt{2}})\rightarrow \theta_1\\
	        E(X, \frac{Z+X}{\sqrt{2}})\rightarrow \phi_2\\
	        E(X, \frac{Z-X}{\sqrt{2}})\rightarrow \theta_2
        \end{split}
    \end{equation*}
    
    Similarly we can collect measurement results $P_2(l_i)$. Now calculate $P(l_i)=\frac{1}{2}[P_1(l_i)+P_1(l_i)]$. We know that the new test data belongs to sublabel $l_m$ where $P(l_m)$ is the maximum.

\end{document}